\documentclass[12pt]{article}
\usepackage{graphicx}
\usepackage{epstopdf}
\usepackage{amsmath}
\usepackage{amsfonts}
\usepackage{amssymb}
\usepackage{latexsym}
\usepackage{hyperref}
\usepackage{bm}
\usepackage{bbold}
\usepackage[english]{babel}
\usepackage[utf8]{inputenc}
\usepackage{xcolor}
\usepackage[colorinlistoftodos]{todonotes}
\usepackage{graphicx}
\usepackage{caption}
\usepackage[caption=false]{subfig}
\usepackage{cite}
\usepackage{float}

\usepackage{setspace}
\allowdisplaybreaks

\usepackage{bigstrut}

\begin{document}

{\onehalfspacing
\begin{center}

{\Large\bf Experimental signatures of Kalb-Ramond-like particles}

\vspace*{0.3cm}

P. C. Malta$^{*}$\\
\vspace*{0.1cm}
{R. Antonio Vieira 23, 22010-100, Rio de Janeiro, Brazil }\\

\vspace*{0.3cm}

J. P. S. Melo$^{**}$\\
\vspace*{0.1cm}
{Departamento de Astrof\'isica, Cosmologia e Intera\c c\~oes Fundamentais (COSMO),\\ Centro Brasileiro de Pesquisas F\'{i}sicas (CBPF), \\Rua Dr Xavier Sigaud 150, Urca, Rio de Janeiro, Brazil, CEP 22290-180 }\\

\vspace*{0.3cm}

C. A. D. Zarro$^{\dagger}$\\
\vspace*{0.1cm}
{Instituto de F\'{\i}sica, Universidade Federal do Rio de Janeiro\\
 Av. Athos da Silveira Ramos, 149
Centro de Tecnologia - bloco A - Cidade Universitária - Rio de Janeiro - RJ - CEP: 21941-909}

\vspace*{0.2cm}
\end{center}

\begin{abstract}
We analyse phenomenological signatures of Kalb-Ramond-like particles, described by an antisymmetric rank-2 tensor, when coupled to fermionic matter. The latter is modelled by a tensor current coupled directly to the Kalb-Ramond field or by a pseudovector current coupled to the rank-1 dual of the field-strength tensor. We obtain limits on the coupling constants to fermions and mass of the Kalb-Ramond-like particles by investigating their impact on the hyperfine splittings of hydrogen, the tree-level unitarity of the $S$ matrix for $e^- + e^+ \rightarrow \ell^- + \ell^+$ scattering with $\ell \neq e$ and the differential cross section for Bhabha scattering. Assuming that the couplings to fermions are independent of the fermion species, the strongest 95\%-CL bounds we find are from LEP data at $\sqrt{s} = 136.23$~GeV, namely $g_{\rm PV}/\Lambda \lesssim 6.3 \times 10^{-13} \, {\rm eV}^{-1}$ and $g_{\rm T} \lesssim 1.3 \times 10^{-12} \left( m/{\rm eV} \right)$, both for $m \ll \sqrt{s}$ (here $\Lambda$ is a characteristic energy scale). These limits can be improved in upcoming lepton colliders due to enhanced precision, as well as higher energies. 

\noindent
  
\end{abstract}}

\vfill
\noindent\underline{\hskip 140pt}\\[4pt]
{$^{*}$ E-mail address: pedrocmalta@gmail.com} \\
\noindent
{$^{**}$ E-mail address: jpsm@cbpf.br} \\
\noindent
{$^{\dagger}$ E-mail address: carlos.zarro@if.ufrj.br}


\section{Introduction}  \label{sec_intro}
\indent

Antisymmetric fields of rank-2 were introduced in 1966 in the context of quantum field theory~\cite{notoph1} (for a review, see Ref.\cite{notoph2}) and eight years later in the dawn of string theory when Kalb and Ramond showed that such fields naturally couple to strings as gauge mediators~\cite{KR}. It was later demonstrated that Kalb-Ramond (KR) fields are necessary to ensure some gauge-invariance properties of a string model describing strong interactions~\cite{Cremmer}. More recently, other aspects of KR fields in string theory, such as in string dualities, have been thoroughly discussed~\cite{Becker}. Stringy KR fields, particularly from low-scale models~\cite{Dvali}, are interesting due to their phenomenological impact. One such case is that of massless KR fields behaving as singlets under the gauge group of the Standard Model (SM) and interacting with fermions via tensor and pseudo-tensor couplings~\cite{Dick0}. Tensor currents interacting directly with KR fields appear in the context of dark matter to generate dipole couplings that strongly suppress the interaction with the visible sector of the SM~\cite{EW_KR, Cline, Dick}. Incidentally, KR fields -- connected to string theories or not -- are used in investigations of confinement~\cite{NambuKR,greensite,Grigorio:2011pi, Guimaraes:2012tx, Smailagic:2020kep}, condensed matter~\cite{Franz:2007}, holography~\cite{Rougemont:2015gia}, the Casimir effect~\cite{Barone:2005hn,Belich:2010xj}, quantum electrodynamics (QED)~\cite{Smailagic:2021poa}, the coupling of KR fields to branes~\cite{Barone:2010zi}, Lorentz-symmetry violation~\cite{Petrov1, Petrov2, Petrov3}, gravitation~\cite{Ferreira:2008yf,Chakraborty:2017uku,Junior:2024ety} and cosmology~\cite{Cosm_1, Cosm_2, Cosm_3}.

Interestingly enough, antisymmetric rank-2 fields are a possible representation of the Lorentz group in four dimensions,  $(1,0)\oplus(0,1)$. This allows a paradigm change to treat the field excitations as particles in their own right: KR-like particles (KRLPs)~\cite{Capanelli:2023uwv}, which may be investigated as new degrees of freedom beyond the SM with phenomenological consequences. Massive KRLPs describe spin-1 particles, analogous -- but not equivalent -- to the de Broglie-Proca field. This suggests a duality~\cite{dual} between the two descriptions: in the massive case, the noninteracting KR field is dual to a pseudovector~\cite{Hell}. In the massless and noninteracting limit, the KR field is dual to a pseudoscalar field, prompting the connection to axion-like particles (ALPs)~\cite{Witten, Russell, PQ, Weinberg, Wilczek}. Similarly, string theory gives rise the so-called hidden photons (HPs), spin-1 particles kinetically mixed with the photon~\cite{Okun, Holdon, Abel1, Abel2, Fayet1}. Just like in the case of de Broglie-Proca, HPs are parity-odd vectors, whereas the KR field is parity-even. For more details on the KR dualities, see Ref.~\cite{Smailagic}. Both ALPs and HPs are well-motivated dark-matter candidates~\cite{DM1, DM2, DM3, DM4, DM5}.

As mentioned above, the dualities of the massless and massive KRL theories suggest a comparison with pseudoscalars and pseudovectors. The most prominent example of pseudoscalars are ALPs, which can have a variety of couplings~\cite{Graham,Capozzi, FIPS}. For example, ALPs may couple to photons via $\mathcal{L}_{\rm int} \supset aF_{\mu\nu} \tilde{F}^{\mu\nu}$, where $\tilde{F}^{\mu\nu}$ is the dual of the electromagnetic field-strength tensor and $a$ is the ALP field. This coupling leads to ALP-photon oscillations, a feature not shared with KRLPs. Electrons and nucleons can also interact with ALPs via $\mathcal{L}_{\rm int} \supset a\bar{\psi} \gamma^5 \psi$ or $\mathcal{L}_{\rm int} \supset \left( \partial_\mu a \right) \bar{\psi} \gamma^\mu \gamma^5 \psi$, the latter being analogous to our pseudovector (PV) coupling, cf. Eq.~\eqref{eq_ints}.

Also relevant are HPs, light spin-1 particles kinetically mixed to photons with a parameter $\chi \ll 1$. Upon diagonalization, the kinetic terms of the HP and the usual photon decouple, but now the electric current couples to the HP with an effective charge $e\chi$, implying that the fermions acquire a ``millicharge" under the new $U_\chi(1)$~\cite{JJ}. Moreover, the mass terms are not diagonal and we have HP-photon oscillations, which is absent from the dynamics of KRLPs, further differentiating the two in their phenomenology. Finally, HPs are true vectors, whereas KRLPs are parity-even tensors, thus causing the respective couplings to matter to differ. There are two natural couplings possible: $\mathcal{L}_{\rm int} \supset e\chi X_\mu \bar{\psi} \gamma^\mu \psi$ and $\mathcal{L}_{\rm int} \supset X_{\mu\nu} \bar{\psi} \sigma^{\mu\nu} \psi$, where $X_\mu$ and $X_{\mu\nu}$ are the HP 4-vector and its field-strength tensor, respectively. The first has no parallel in the case of KRLPs because of the parity properties of the fields involved. The second is similar to the tensor (T) interaction, which will be investigated in this work, cf. Eq.~\eqref{eq_ints}. Note that for HPs, this coupling is actually derivative because of the presence of the HP field-strength tensor, but in the case of KRLPs it is the field itself that couples to the T current.

Here we confine ourselves to massive interacting KRLPs, not distinguishing among the possible underlying ultraviolet theories. Similarly to HPs, KRLPs may be broadly described as generic Z' bosons feebly interacting with SM particles~\cite{Leike, Langacker}. However, contrary to HPs -- or even axions or ALPs -- we do not consider a direct coupling of the KRLPs to photons, only to spin-1/2 matter. The aforementioned PV and T couplings to fermions would give rise to modifications of well-measured quantities, such as spectral lines and differential cross sections. Here we compare the predictions including KRLP-mediated interactions with experimental data to set exclusion limits on the PV and T couplings, as well as on the KRLP mass.

This paper is organized as follows: in Sec.~\ref{sec_theory} we set up the basics of the theory of KRLPs and introduce the interaction to matter. Next, in Sec.~\ref{sec_bounds} we investigate possible phenomenological signatures of KRLPs. In  Sec.~\ref{sec_H_atom} we obtain limits from the hyperfine structure of hydrogen, whereas in Sec.~\ref{sec_unitarity} we study the tree-level unitarity bounds for $e^- + e^+ \rightarrow \ell^- + \ell^+$ ($\ell \neq e$) and in  Sec.~\ref{sec_bhabha} we analyse Bhabha scattering in lepton colliders. Finally, in Sec.~\ref{sec_conclusions} we present our conclusions. The flat metric is $\eta_{\mu\nu} = {\rm diag}(+,-,-,-)$ and the totally antisymmetric Levi-Civita symbol $\epsilon^{\mu\nu\lambda\kappa}$ is defined such that $\epsilon^{0123} = +1$ with $\epsilon^{0ijk} = -\epsilon_{0ijk} = \epsilon^{ijk} = \epsilon_{ijk}$ (Greek indices run from 0 to 3; Latin indices run from 1 to 3). Four-vectors are defined such that $V^\mu = (V^0, {\bf V})$ and $V_\mu = (V_0, -{\bf V})$ with $V_0 = V^0$; the single components of ${\bf V}$ are denoted by ${\bf V}_k$. We use natural units ($\hbar = c = 1$) throughout. The calculations of Sec.~\ref{sec_bhabha} were cross checked with the FeynCalc Mathematica Package~\cite{feyncalc1,feyncalc2,feyncalc3,feyncalc4}.


\section{Kalb-Ramond-like particles and their interactions}  \label{sec_theory}
\indent

In order to determine what kind of experimental signatures a KRLP could produce as a kind of Z', it is important to discuss its properties, as well as those from the fermion currents with which it could couple.

The dynamics of the real antisymmetric rank-2 KR field $B_{\mu\nu}$ is described by the Lorentz-invariant action
\begin{equation} \label{eq_S_KR}
S = \int d^4x \left[ \frac{1}{6}H_{\mu\nu\lambda} H^{\mu\nu\lambda} - \frac{1}{2}m^2 B_{\mu\nu} B^{\mu\nu} + \mathcal{L}_{\rm int} \right] \, ,
\end{equation}
where the rank-3 field-strength tensor is $H_{\mu\nu\lambda} = \partial_\mu B_{\nu\lambda} + \partial_\nu B_{\lambda\mu} + \partial_\lambda B_{\mu\nu}$ and $m > 0$ is the mass of the KRLP. The KR field has canonical mass dimension one, whereas its field-strength tensor has dimension two. Here $\mathcal{L}_{\rm int}$ describes the interaction with matter, to be discussed below. In the massless theory, the field-strength tensor allows for a gauge symmetry, namely, $B_{\mu\nu} \rightarrow B_{\mu\nu} + \partial_\mu \Omega_\nu - \partial_\nu \Omega_\mu$, with the 4-vector $\Omega_\mu(x)$ being an arbitrary function. Note that, besides the aforementioned gauge transformation, $B_{\mu\nu}$ is itself left unchanged under $\Omega_\mu \rightarrow \Omega_\mu + \partial_\mu \lambda$ for a generic $\lambda(x)$. For $m \neq 0$ the model is not gauge invariant -- this is the paradigm assumed from here on.

Let us first focus on the case without interactions with equations of motion 
\begin{equation} \label{eq_motion}
\partial_\mu H^{\mu\nu\lambda} - m^2 B^{\nu\lambda} = 0 \, .
\end{equation}
Due to its antisymmetry, $B_{\mu\nu}$ has six independent components. However, the subsidiary condition $\partial_\mu B^{\mu\nu} = 0$ contributes with three linearly independent equations, reducing the number of degrees of freedom to three: KRLPs are neutral, parity-even, massive spin-1 particles.

The fundamental field is the parity-even $B_{\mu\nu}$, to which a parity-odd field-strength tensor $H^{\mu\nu\lambda}$ is associated. We can also define its dual, $\tilde{H}_\mu = \frac{1}{6} \epsilon_{\mu\nu\rho\sigma} H^{\nu\rho\sigma}$, which is a pseudovector. The fermionic currents coupling to either $B_{\mu\nu}$ or $\tilde{H}_\mu$ are of two distinct types: pseudovector (PV) and tensor (T), given by $j^{\mu}_{\rm PV} = \bar{\psi} \gamma^\mu \gamma^5 \psi$ and $j^{\mu \nu}_{\rm T} = \bar{\psi} \sigma^{\mu\nu} \psi$, respectively, with $\sigma^{\mu\nu} = \frac{i}{2}\left[ \gamma^\mu, \gamma^\nu \right]$. Here $\psi$ denotes a massive spin-1/2 fermion $f$. The PV current couples to $\tilde{H}_\mu$, whereas the T current couples directly to the KRL field, $B_{\mu\nu}$, so the natural choices for $\mathcal{L}_{\rm int}$ are  
\begin{equation} \label{eq_ints}
\mathcal{L}_{\rm PV} = \frac{g_{\rm PV}^f}{\Lambda} \tilde{H}_\mu \bar{\psi} \gamma^\mu \gamma^5  \psi \quad \text{or} \quad \mathcal{L}_{\rm T} = g_{\rm T}^f B_{\mu\nu} \bar{\psi} \sigma^{\mu\nu} \psi \, .
\end{equation} 
The real and dimensionless coupling constants $g_{\rm PV, T}^f$ may depend on the fermion species~$f$; we only consider vertices connecting fermions of the same family. The real parameter $\Lambda$ has dimension of mass and is related to the energy scale of the underlying physics. These are free parameters that, together with $m$, form the parameter space of the theory.

In QED the vector current $j^\mu_{\rm V} = \bar{\psi} \gamma^\mu \psi$ is conserved due the local $U(1)_{\rm em}$ symmetry, a remanent of the $SU(2)_{\rm L} \otimes U(1)_{\rm Y}$ electroweak symmetry. Other, less trivial, current structures $j = \bar{\psi} \Gamma \psi$ with $\Gamma = \{ \gamma_\mu \gamma_5, \sigma_{\mu\nu}, ... \}$ are nonetheless possible. Pseudovector currents naturally arise in weak interactions due to the embedded V-A structure~\cite{Nambu}, but also in axion-nucleon effective interactions~\cite{Cao}. Tensor currents, for example, are necessary to describe mesons in the Nambu–Jona-Lasinio model~\cite{Naydenov}, also if one includes dark vector bosons~\cite{tensor1}. Similarly, tensor currents have been proposed to explain the discrepancies between theory and experiment concerning the anomalous magnetic moment of the muon with massive dark bosons~\cite{tensor2}, as fundamental ingredients of dimension-six operators involving a novel massless spin-1 boson~\cite{Dobrescu}.

Current conservation translates to $\partial\cdot j= 0$, where the current is either the vector current $j^\mu_{\rm V} = \bar{\psi} \gamma^\mu \psi$ of electrodynamics, the PV current $j^{\mu}_{\rm PV} = \bar{\psi} \gamma^\mu \gamma^5 \psi$ or the T current $j^{\mu \nu}_{\rm T} = \bar{\psi} \sigma^{\mu\nu} \psi$. The vector current is conserved, since the Lagrangian is invariant under $U(1)_{\rm em}$. In fact, by using the Dirac equation we can show that 
\begin{eqnarray} 
\partial_\mu j^{\mu}_{\rm PV} & = & 2im_f \bar{\psi} \gamma^5 \psi  \, , \label{eq_dj_pv} \\
\partial_\mu j^{\mu\nu}_{\rm T} & = &  -2m_f j^\nu_{\rm V}  + i \left[ \bar{\psi} \left( \partial^\nu  \psi \right) - \left( \partial^\nu  \bar{\psi} \right)  \psi \right] \, . \label{eq_dj_t} 
\end{eqnarray}
The PV current is only conserved in the case of massless fermions; the T current, on the other hand, is not. For $m = 0$, the KRLP Lagrangian exhibits gauge symmetry (see comment below Eq.~\eqref{eq_S_KR}) and a tensor current would be conserved~\cite{Bailey}. In the massive case, however, these currents are not conserved, meaning that the momentum-dependent contributions from the longitudinal parts of the propagator will not automatically vanish.

From Eq.~\eqref{eq_ints} we read off the Feynman rule for the T vertex: 
\begin{equation} \label{eq_vertex_T}
V_{\rm T}^{\mu\nu} = ig_{\rm T}^f \sigma^{\mu\nu}  \, .
\end{equation}
The PV vertex is more involved. Since $B_{\mu\nu}$ is the fundamental field, this interaction is derivative, thus introducing factors of the 4-momentum carried by the KRLP. Using $\partial_\mu \rightarrow -ik_\mu$ with the 4-momenta flowing into the vertex defined as positive, the PV interaction vertex is
\begin{equation}\label{eq_vertex_PV}
V_{\rm PV}^{\alpha\beta} = \frac{g_{\rm PV}^f}{4\Lambda}\epsilon^{\mu\nu\alpha\beta} \left( k_\nu \gamma_\mu - k_\mu \gamma_\nu  \right) \gamma^5 \, .
\end{equation}

%

The propagator may be obtained by writing the free KRLP action in bilinear form as $S = \frac{1}{2} \int d^4x B^{\nu\lambda} \mathcal{O}_{\nu\lambda,\alpha\beta}(x) B^{\alpha\beta}$. In momentum space we have
\begin{eqnarray} \label{eq_prop_1}
\mathcal{O}^{\nu\lambda,\alpha\beta}(k) & = & \frac{1}{2} \Big[  \left( \eta^{\nu\alpha} \eta^{\lambda\beta} - \eta^{\lambda\alpha} \eta^{\nu\beta} \right) (k^2 - m^2) \nonumber \\
& + & \left( \eta^{\nu\beta}k^\lambda - \eta^{\lambda\beta}k^\nu \right)k^\alpha + \left( \eta^{\lambda\alpha}k^\nu - \eta^{\nu\alpha}k^\lambda \right)k^\beta \Big] \, ,
\end{eqnarray}
which is antisymmetric in $(\nu\lambda)$ and $(\alpha\beta)$. Note the symmetry in the exchange $(\nu\lambda) \leftrightarrow (\alpha\beta)$. The propagator is defined as the inverse of this operator, $\Delta(k) = i \mathcal{O}^{-1}(k)$, and is calculated in App.~\ref{app_A}; here we quote the result with adequate indices for convenience
\begin{equation} \label{eq_prop_final_2}
\Delta_{\nu\lambda,\alpha\beta}(k) = \frac{i}{k^2 - m^2}\left[ \left( 1^{a.s.} \right)_{\nu\lambda,\alpha\beta} -  \frac{k^2}{m^2}\left( P^1_e \right)_{\nu\lambda,\alpha\beta} \right] \, ,
\end{equation}
where $\left( 1^{a.s.} \right)_{\nu\lambda,\alpha\beta}$ is the antisymmetric identity and $P^1_e$ is one of the projection operators for rank-2 tensors, both defined in App.~\ref{app_A}. The presence of the pole at $k^2 = m^2$ confirms that the KRLP is indeed massive.

\section{Phenomenology}  \label{sec_bounds}
\indent

Having developed the basics of KRLPs as an alternative description of massive spin-1 particles, now we analyse their phenomenology. To this end we use experimental data from different sources, in particular, spectroscopy and scattering at high energies. These are chosen, respectively, due to their high precision and high energies attained. We also study perturbative unitarity to establish a consistent scale for the KRLP couplings to fermions and its mass.

\subsection{Atomic physics}  \label{sec_H_atom}
\indent

In order to calculate the shift in the spectral lines caused by the exchange of a KRLP, we must first determine the interaction potential between spin-1/2 point sources. The positive-energy solution of the Dirac equation for a fermion $f$ with mass $m_f$, spin $s$ and 4-momentum $p^\mu = \left( E, {\bf p} \right)$ satisfying $E^2 - {\bf p}^2 = m_f^2$ is
\begin{equation}\label{eq_spinor}
u_s(p) = N_{\rm R} \left( \begin{array}{c}
\xi_s \\\frac{{\bm \sigma}\cdot {\bf p}}{E+m_f} \xi_s  \end{array} \right) 
\end{equation}
with ${\bm \sigma}$ the Pauli spin-matrix vector and $N_{\rm R} = \sqrt{E+m_f}$ being the relativistic normalization; in a non-relativistic (NR) treatment we change it to $N_{\rm R} \rightarrow N_{\rm NR} = N_{\rm R}/\sqrt{2E}$~\cite{Maggiore}. The spinor normalization is $u_s^\dagger(p) u_{s'}(p) = 2E \delta_{ss'}$ and here we will ignore spin flip, so $s = s'$.

We are interested in the tree-level interaction of fermion $a$ with initial and final 4-momenta $p_1$ and $p_3$, respectively, with fermion $b$, whose initial and final 4-momenta are $p_2$ and $p_4$, respectively. In this case only one Feynman diagram -- the $t$-channel diagram, analogous to those in Fig.~\ref{fig_diagrams} -- contributes and the amplitude reads (suppressing spin indices for clarity)
\begin{equation} \label{eq_amp_H_1}
i\mathcal{M} = \bar{u}(p_3) V^{\mu\nu}_v u(p_1) \Delta_{\mu\nu,\alpha\beta}(q) \bar{u}(p_4) V^{\alpha\beta}_v u(p_2)  \, ,
\end{equation}
where the propagator of the KRLP is given by Eq.~\eqref{eq_prop_final_2} with $q = p_3 - p_1 = p_2 - p_4$. Here $v = \{ {\rm PV,T} \}$ indicates the specific interaction being considered; we only consider the potentials generated by vertices of the same type, that is, the vertices are either both PV or T, but not a mixture (see Refs.~\cite{Grupo1, Grupo2} for a discussion). More explicitly, the relativistic PV and T amplitudes are
\begin{eqnarray} \label{eq_amp_PV_atom}
\mathcal{M}_{\rm PV} & = & -\dfrac{ g_{\rm PV}^{a} g_{\rm PV}^{b}  }{2\Lambda^2 (q^2 -m^2)} \bigg\{ q^2 \Big[\bar{u}^{(s_3)}(p_3)  \gamma_\mu \gamma_5 u^{(s_1)}(p_1) \Big]  \Big[\bar{v}^{(s_2)}(p_2)  \gamma^\mu  \gamma_5 v^{(s_4)}(p_4) \Big]  \nonumber \\
 & - & q_\mu q_\nu \Big[\bar{u}^{(s_3)}(p_3)  \gamma^\mu \gamma_5 u^{(s_1)}(p_1) \Big]  \Big[\bar{v}^{(s_2)}(p_2)  \gamma^\nu  \gamma_5 v^{(s_4)}(p_4) \Big]  \bigg\}
\end{eqnarray}
and
\begin{eqnarray} \label{eq_amp_T_atom}
\mathcal{M}_{\rm T} & = & - \dfrac{ g_{\rm T}^{a} g_{\rm T}^{b} }{(q^2 -m^2)}\bigg\{   \Big[\bar{u}^{(s_3)}(p_3) \sigma_{\mu\nu} u^{(s_1)}(p_1) \Big]  \Big[\bar{v}^{(s_2)}(p_2) \sigma^{\mu \nu} v^{(s_4)}(p_4) \Big]  \nonumber \\
& + & \frac{2q^\nu q_\alpha}{m^2} \Big[\bar{u}^{(s_3)}(p_3) \sigma_{\mu\nu} u^{(s_1)}(p_1) \Big]  \Big[\bar{v}^{(s_2)}(p_2) \sigma^{\mu \alpha} v^{(s_4)}(p_4) \Big]  \bigg\} \, .
\end{eqnarray}

From scattering theory, the interaction potential is related to the amplitude (at leading order) by the first Born approximation~\cite{Maggiore}
\begin{equation} \label{eq_def_pot}
V(r) = - \int \frac{d^3{\bf q}}{\left( 2\pi \right)^3} \mathcal{M}_{\rm NR}({\bf q}) \, e^{i {\bf q}\cdot{\bf r}} \, ,
\end{equation}
where ${\bf q}$ is the 3-momentum carried by the mediator and $r = |{\bf r}|$ is the distance between the sources. We must now consider the NR limit of the amplitudes~\eqref{eq_amp_PV_atom} and~\eqref{eq_amp_T_atom}. For this we use the spinor~\eqref{eq_spinor} with $N_{\rm R} \rightarrow N_{\rm NR} = N_{\rm R}/\sqrt{2E}$, as well as the approximation of an elastic scattering, to evaluate the bilinears. For further details, please refer to App.~\ref{app_B}. The results are the NR amplitudes for PV and T sources, cf. Eqs.~\eqref{eq_app_amp_PV_final} and~\eqref{eq_app_amp_T_final}, which can be used to compute the respective potentials with the help of the Fourier integrals listed in App.~\ref{app_ints}. The ensuing parity-even potentials due to the exchange of a KRLP between fermions $a$ and $b$ in the NR limit are
\begin{eqnarray}
V_{\rm PV}(r) & = & \left( \frac{ g_{\rm PV}^a g_{\rm PV}^b }{2 \Lambda^2} \right) \Bigg\{ \left[ \frac{2}{3}\delta^3({\bf r}) - \left( 1 + mr + m^2 r^2 \right) \frac{e^{-mr}}{4\pi r^3} \right] {\bm \sigma}_a \cdot {\bm \sigma}_b  \nonumber \\
& & \quad \quad \quad \quad  \quad \quad \quad   +\left( 3 + 3mr + m^2 r^2 \right) \frac{e^{-mr}}{4\pi r^3} \left( \hat{ {\bf r}} \cdot {\bm \sigma}_a \right) \left( \hat{ {\bf r}} \cdot {\bm \sigma}_b \right) \Bigg\}   \, , \label{eq_pot_PV_KRLP} \\
V_{\rm T}(r) & = & -\left( \frac{2g_{\rm T}^a g_{\rm T}^b}{m^2} \right)  \Bigg\{ \left[ \frac{2}{3}\delta^3({\bf r}) - \left( 1 + mr \right) \frac{e^{-mr}}{4\pi r^3} \right] {\bm \sigma}_a \cdot  {\bm \sigma}_b  \nonumber \\
& & \quad \quad \quad \quad  \quad \quad \quad   +\left( 3 + 3mr + m^2 r^2 \right) \frac{e^{-mr}}{4\pi r^3} \left( \hat{ {\bf r}} \cdot {\bm \sigma}_a \right) \left( \hat{ {\bf r}} \cdot {\bm \sigma}_b \right) \Bigg\}   \, . \label{eq_pot_T_KRLP} 
\end{eqnarray}

A few comments are in order. The heaviest fermion in our analysis is the proton, the electron is the lightest. Even though we make no assumption regarding the scale of $m$, it is possible that the KRLP is lighter than either of the fermions (but not strictly massless). We may thus generally neglect terms $\mathcal{O}\left( |{\bf P}|^2/m_a m_b \right)$ in the amplitudes. From the decomposition of the fermionic bilinears, cf. App.~\ref{app_B_amps}, and the structure of the propagator~\eqref{eq_prop_final_2} we see that the $0i$ and $ij$ components do not mix in the NR amplitudes, causing the only momentum contributions to be automatically of second order. The typical velocity of the electron in the hydrogen atom is $\approx \alpha \approx 1/137$, so all things being equal, the velocity-dependent terms are suppressed by a factor $\approx 5 \times 10^{-5}$. We therefore work in the static limit.

The presence of the spins of both fermions is also relevant, though this is not new in the literature of exotic potentials~\cite{Grupo1, Grupo2, Grupo3, Fadeev, Fadeev2, Cong, Cong2, Dobrescu2,Joao}. It shows that the energy shifts resulting from using the PV and T potentials as perturbations will have a structure similar to that of the hyperfine interaction~\cite{Cohen_book}. Moreover, both potentials have contact terms $\sim \delta^3({\bf r})$, also featured in other potentials involving the exchange of spin-1 mediators~\cite{Grupo1, Grupo2, Grupo3, Fadeev, Fadeev2, Cong, Cong2}. These terms are not important in interactions at macroscopic distances, but they do play a role in atomic phenomena, specially when one considers wave functions that are finite at the origin~\cite{Fadeev2, Cohen_book}.

The limit of a light KRLP is noteworthy. The PV potential is unproblematic, but the T potential is enhanced due to the pre-factor $1/m^2$. A similar feature appears in the context of the Proca-mediated interaction between spin-1/2 fermions with PV couplings $\sim V_\mu \bar{\psi} \gamma^\mu \gamma^5 \psi$, where $V_\mu$ is the massive mediator~\cite{Grupo2}. This kind of PV coupling is also considered in Refs.~\cite{Fadeev, Fadeev2}, with similar conclusions. Note that this PV interaction differs from the one we consider, cf. Eq.~\eqref{eq_ints}, where the dual field-strength tensor $\tilde{H}_\mu$ couples to the PV current, not to $B_{\mu\nu}$ itself, thus bringing extra factors of the momentum transfer, ${\bf q}$, which eventually lead to the cancelling of the overall $1/m^2$ factor present in the NR amplitude. Interestingly, the vertex for T couplings does not carry momentum factors~\cite{Capanelli:2023uwv}; here the coupling of the T current is directly to $B_{\mu\nu}$ and the $1/m^2$ factor survives. For a truly massless mediator, the divergence caused by $\sim q_\mu q_\nu/m^2$ does not occur, since the mediator has no longitudinal degree of freedom~\cite{Fadeev2}.

\subsubsection{Limits from the hyperfine structure of hydrogen} \label{sec_hyperfine_shifts}
\indent

The potentials calculated above may be understood as quantum-mechanical operators acting as small corrections to the unperturbed Hamiltonian of hydrogen (we make $a,b \rightarrow p,e$). The potentials~\eqref{eq_pot_PV_KRLP} and~\eqref{eq_pot_T_KRLP} will then cause certain energy levels to slightly shift and split, thereby inducing the transition frequency between said levels to deviate from those experimentally observed and theoretically calculated using QED. Here we obtain these energy differences between the split levels -- the so-called hyperfine splitting intervals -- of the $n = 1$ and $n =2$ states of hydrogen with $\ell = 0$, that is, the $1s_{1/2}$ and $2s_{1/2}$ states.

In the standard treatment of the hydrogen atom the spin of the electron first appears in the context of the spin-orbit interaction caused by the coupling of the electron's magnetic moment to its orbital motion -- these corrections are of order $\approx 10^{-4}$~eV, marking the energy scale of the fine structure. The spin of the proton enters the picture when we consider the direct coupling of its magnetic moment to that of the electron~\cite{Cohen_book, Griffiths_hyp}: this hyperfine interaction breaks the degeneracy of the $\ell = 0$, spherically symmetric $1s_{1/2}$ and $2s_{1/2}$ states. The associated hyperfine energy splitting is of order $\approx 10^{-6}$~eV for the $1s_{1/2}$ level, which is broken in (three-fold degenerate) triplet and singlet energy states~\cite{Cohen_book, Griffiths_hyp, Parthey, Parthey2, Bullis_hfs}.

The calculation of the splitting caused by the PV and T potentials closely resembles that of the standard hyperfine structure~\cite{Cohen_book}. To start with, we must calculate the matrix elements of the potentials between the states $| \psi_{n} \rangle = | \psi_{n00} \rangle \otimes | m_{z,e} m_{z,p} \rangle$ with $n = 1,2$; here $m_{z, (e,p)} = \pm 1/2$ are the projections of the respective spins, ${\bf S}_{e,p} = {\bm \sigma}_{e,p}/2$, along the quantization axis. The normalized unperturbed wave functions for the $1s_{1/2}$ and $2s_{1/2}$ states of hydrogen are
\begin{eqnarray} 
\langle {\rm r}| \psi_{100} \rangle & = & \psi_{100}(r) = \sqrt{\frac{k^3}{\pi}} e^{-kr} \, , \label{eq_psi_1} \\
\langle {\rm r}| \psi_{200} \rangle & = & \psi_{200}(r) = \frac{1}{2} \sqrt{ \frac{k^3}{2\pi} } \left( 1 - \frac{kr}{2}  \right) e^{-kr/2}  \, , \label{eq_psi_2}
\end{eqnarray}
with $k = 1/a_0$, where $a_0 = 2.68 \times 10^{-4} \, {\rm eV}^{-1}$ is the Bohr radius ($k = 3.73 \times 10^{3}$~eV).

Let us work out the case of the PV potential. Setting ${\bm \sigma} = 2 {\bf S}$, the matrix elements are
\begin{eqnarray}
\langle \psi_n | V_{\rm PV} | \psi_n \rangle & = & C_{\rm PV} \langle \psi_n | \Bigg\{ \left[ \frac{8\pi}{3}\delta^3({\bf r}) - F_1(r) \right] {\bf S}_e \cdot {\bf S}_p  + F_2(r) \left( \hat{ {\bf r}} \cdot  {\bf S}_e \right) \left( \hat{ {\bf r}} \cdot {\bf S}_p \right) \Bigg\} | \psi_n \rangle  \, ,  \label{eq_matrix_element_1} 
\end{eqnarray}
where $C_{\rm PV} = g_{\rm PV}^{e} g_{\rm PV}^{p} / 2\pi\Lambda^2$ and we defined the functions $F_1(r) = \left( 1 + mr + m^2 r^2 \right) e^{-mr}/r^3$ and $F_2(r) = \left( 3 + 3mr + m^2 r^2 \right) e^{-mr}/r^3$. Writing $\langle \psi_n | V_{\rm PV} | \psi_n \rangle = C_{\rm PV} \left( A_c - A_1 + A_2 \right)$, we find
\begin{eqnarray} 
A_c & = & \frac{8\pi}{3} \langle {\bf S}_e \cdot  {\bf S}_p \rangle |\psi_{n00}(0)|^2 \, , \label{eq_A_c} \\
A_1 & = & 4\pi \langle {\bf S}_e \cdot  {\bf S}_p \rangle \int dr \,  r^2 \,  F_1(r) |\psi_{n00}(r)|^2 \, , \label{eq_A_1} \\
A_2 & = & \langle \left({\bf S}_e\right)_i \cdot  \left({\bf S}_p\right)_j \rangle \int d\Omega \left( \hat{ {\bf r}}_i \hat{ {\bf r}}_j \right) \int dr \, r^2 \,  F_2(r) |\psi_{n00}(r)|^2 \, , \nonumber \\
& = & \frac{4\pi}{3} \langle {\bf S}_e \cdot  {\bf S}_p \rangle \int dr \, r^2 \, F_2(r) |\psi_{n00}(r)|^2 \,  \label{eq_A_2}
\end{eqnarray}
where, in the last line, we used $\int d\Omega \left( \hat{ {\bf r}}_i \hat{ {\bf r}}_j \right) = 4\pi \delta_{ij}/3$ due to spherical symmetry. Putting it all together we get
\begin{eqnarray} \label{eq_matrix_element_2}
\langle \psi_n | V_{\rm PV} | \psi_n \rangle & = & \frac{8\pi}{3} C_{\rm PV} \langle {\bf S}_e \cdot  {\bf S}_p \rangle \Bigg\{  |\psi_{n00}(0)|^2 + \frac{1}{2} \int_0^\infty dr \, r^2 \,  \left[ F_2(r) - 3F_1(r) \right]  |\psi_{n00}(r)|^2  \Bigg\} \, , \nonumber \\
& = & \frac{8\pi}{3} C_{\rm PV} \langle {\bf S}_e \cdot  {\bf S}_p \rangle \Bigg\{  |\psi_{n00}(0)|^2 - m^2 \int_0^\infty dr \, r \,  e^{-mr}  |\psi_{n00}(r)|^2  \Bigg\} \, .
\end{eqnarray}

Let us now address the factor  $\langle {\bf S}_e \cdot  {\bf S}_p \rangle = \langle m_{z,e} m_{z,p} | {\bf S}_e \cdot  {\bf S}_p | m_{z,e} m_{z,p} \rangle$.
The total angular momentum of the system is ${\bf F} = {\bf S}_e + {\bf S}_p$, since we are dealing with zero orbital angular momentum. Both the electron and the proton have spin 1/2, whose addition can only generate ${\bf F}$ with eigenvalues $F = 0, 1$: the four eigenstates of the perturbation will have only two distinct energy levels. Noting that ${\bf S}_e \cdot  {\bf S}_p = \left(   {\bf F}^2 -  {\bf S}_e^2 -  {\bf S}_p^2  \right)/2 = \left(   {\bf F}^2 -  3/2  \right)/2$ we may diagonalize this operator to find that its eigenvalues are $\lambda_F = \left[ F(F+1) - 3/2  \right]/2$, so that $\lambda_{F} = -3/4, +1/4$ for $F=0,1$, respectively. We are interested in corrections to $\Delta E_{\rm hfs} (ns)$, which is the energy difference between the triplet ($F = 1$) and singlet ($F = 0$) states for the $n = 1,2$ levels. Since $\lambda_{F=1} - \lambda_{F=0} =1/4 - (-3/4) = 1$, the PV splitting interval is
\begin{equation} \label{eq_delta_hfs_PV}
\Delta E_{\rm hfs}^{\rm (PV)} (ns) = \frac{8\pi}{3} C_{\rm PV} \Bigg\{  |\psi_{n00}(0)|^2 - m^2 \int_0^\infty dr \, r \,  e^{-mr}  |\psi_{n00}(r)|^2  \Bigg\} \, .
\end{equation}
Using the normalized wave functions~\eqref{eq_psi_1} and~\eqref{eq_psi_2} we obtain
\begin{eqnarray}
\Delta E_{\rm hfs}^{\rm (PV)} (1s) & = & \frac{4g_{\rm PV}^{e} g_{\rm PV}^{p}}{3\pi \Lambda^2} k^3 \left[ 1 - \frac{m^2}{\left( m + 2k \right)^2}  \right]     \, , \label{eq_Delta_hfs_PV_1s} \\
\Delta E_{\rm hfs}^{\rm (PV)} (2s) & = & \frac{g_{\rm PV}^{e} g_{\rm PV}^{p}}{6\pi \Lambda^2} k^3 \left[ 1 - \frac{m^2 \left( k^2 + 2m^2 \right)}{2\left( m + k \right)^4}  \right]  \, . \label{eq_Delta_hfs_PV_2s} 
\end{eqnarray}
Note that in the massless limit we have $\Delta E_{\rm hfs}^{\rm (PV)} (1s)/\Delta E_{\rm hfs}^{\rm (PV)} (2s) = 8$.

The steps above may be repeated for the T potential, cf. Eq.~\eqref{eq_pot_T_KRLP}, but here we must make the substitution $F_1(r) \rightarrow F_3(r) = \left( 1 + mr \right) e^{-mr}/r^3$. This leads to
\begin{eqnarray} \label{eq_matrix_element_2}
\langle \psi_n | V_{\rm T} | \psi_n \rangle & = & -\frac{8\pi}{3} C_{\rm T} \langle {\bf S}_e \cdot  {\bf S}_p \rangle \Bigg\{  |\psi_{n00}(0)|^2 + \frac{1}{2} \int_0^\infty dr \, r^2 \,  \left[ F_2(r) - 3F_3(r) \right]  |\psi_{n00}(r)|^2  \Bigg\} \, , \nonumber \\
& = & -\frac{8\pi}{3} C_{\rm T} \langle {\bf S}_e \cdot  {\bf S}_p \rangle \Bigg\{  |\psi_{n00}(0)|^2 + \frac{m^2}{2} \int_0^\infty dr \, r \,  e^{-mr}  |\psi_{n00}(r)|^2  \Bigg\} \, 
\end{eqnarray}
with $C_{\rm T} = 2g_{\rm T}^{e^-}  g_{\rm T}^{p^+}  / \pi m^2$. The hyperfine splitting interval due to the T potential is then
\begin{equation} \label{eq_delta_hfs_T}
\Delta E_{\rm hfs}^{\rm (T)} (ns) = -\frac{8\pi}{3} C_{\rm T} \Bigg\{  |\psi_{n00}(0)|^2 + \frac{m^2}{2} \int_0^\infty dr \, r \,  e^{-mr}  |\psi_{n00}(r)|^2  \Bigg\} \, ,
\end{equation}
which for $n=1,2$ is explicitly given by
\begin{eqnarray}
\Delta E_{\rm hfs}^{\rm (T)} (1s) & = & -\frac{16 g_{\rm T}^{e} g_{\rm T}^{p}}{3\pi m^2} k^3 \left[ 1 + \frac{m^2}{2\left( m + 2k \right)^2}  \right]  \, , \label{eq_Delta_hfs_T_1s} \\
\Delta E_{\rm hfs}^{\rm (T)} (2s) & = & -\frac{2g_{\rm T}^{e} g_{\rm T}^{p}}{3\pi m^2} k^3 \left[ 1 + \frac{m^2 \left( k^2 + 2m^2 \right)}{4\left( m + k \right)^4}  \right]  \, . \label{eq_Delta_hfs_T_2s} 
\end{eqnarray}

The hyperfine structure is caused by finite wave functions at $r=0$ and is sensitive to the finite size of the proton. The precision of theoretical calculations is dominated by uncertainties in the modelling of the proton's internal structure. It is thus convenient to consider Sternheim’s splitting interval $D_{21} = 8 \Delta E_{\rm hfs}(2s) - \Delta E_{\rm hfs}(1s)$~\cite{D21}, which reduces the uncertainty from nuclear-structure effects since $|\psi_{100}(0)|^2 = 8|\psi_{200}(0)|^2$, cf. Eqs.~\eqref{eq_psi_1} and~\eqref{eq_psi_2}. The results above give
\begin{eqnarray}
D_{21}^{\rm (PV)} & = & \frac{4g_{\rm PV}^{e} g_{\rm PV}^{p}}{3\pi \Lambda^2} m^2 k^3 \left[ \frac{1}{\left( m + 2k \right)^2} - \frac{\left( k^2 + 2m^2 \right)}{2\left( m + k \right)^4}   \right]      \, , \label{eq_D_21_PV} \\
D_{21}^{\rm (T)} & = &  \frac{4g_{\rm T}^{e} g_{\rm T}^{p}}{3\pi} k^3 \left[ \frac{2}{\left( m + 2k \right)^2}  - \frac{\left( k^2 + 2m^2 \right)}{\left( m + k \right)^4}  \right]   \, . \label{eq_D_21_T}
\end{eqnarray}

In order to constrain the unknown KRLP parameters we must compare our predictions for $D_{21}$ to a reference value $\Delta D$. Following Refs.~\cite{Fadeev, Fadeev2, Ficek, Cong} we obtain $\Delta D$ by evaluating 
\begin{equation} \label{eq_int_bound}
\int_{-\Delta D}^{\Delta D} \frac{1}{\sqrt{2\pi} \sigma} e^{-(x - \mu)^2/(2\sigma^2)} dx = f(\Delta D) \, .
\end{equation}
Here $\sigma = \sqrt{\sigma_{\rm th}^2 + \sigma_{\rm exp}^2}$ is the combined error from theory and experiment and $\mu$ is the (mean) deviation of experiment from the theoretical prediction. Choosing $f(\Delta D) = 0.95$ gives a $95\%$ confidence-level (CL) limit. The meaning of Eq.~\eqref{eq_int_bound} is simple: the best theoretical calculations are statistically compatible with the best measurements of $D_{21}$. Therefore, our contributions to $D_{21}$, cf. Eqs.~\eqref{eq_D_21_PV} and~\eqref{eq_D_21_T}, cannot be arbitrarily large and $\Delta D$ is the largest deviation from the mean $\mu$ which still preserves the agreement between experiment and standard theory.

\begin{figure}[t!]
\begin{minipage}[b]{1.0\linewidth}
\includegraphics[width=\textwidth]{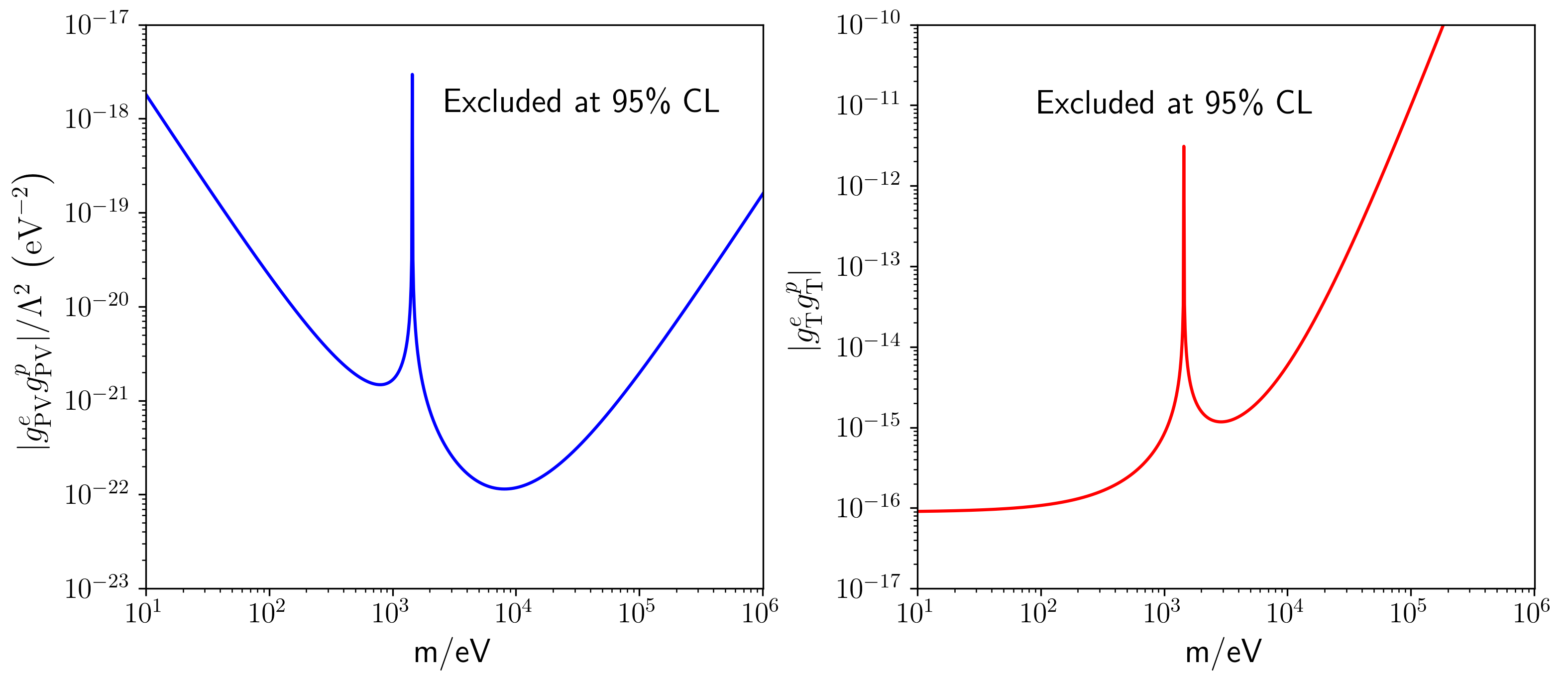}
\end{minipage} \hfill
\caption{ Constraints on the parameter space of KRLPs coupled to electrons and protons at $95\%$~CL extracted by comparing experimental and theoretical values for Steinheim's splitting interval $D_{21} = 8 \Delta E_{\rm hfs}(2s) - \Delta E_{\rm hfs}(1s)$ in hydrogen.  }
\label{fig_exc_hfs}
\end{figure}

The most recent measurement of the hyperfine splitting of the $2s_{1/2}$ level in hydrogen gives $D_{21}^{\rm exp} = 48.9592(68)$~kHz~\cite{Bullis_hfs}, whereas the theoretical prediction is $D_{21}^{\rm th} = 48.9541(23)$~kHz~\cite{D21_th}. Here we merely quote the transition frequencies without explicitly converting them to energies. We have $\mu = D_{21}^{\rm th} - D_{21}^{\rm exp} = -0.0051$~kHz with $\sigma = 0.0072$~kHz~\cite{Cong}; plugging these values in Eq.~\eqref{eq_int_bound} yields $\Delta D = 0.0170 \, {\rm kHz} = 7.04 \times 10^{-14}$~eV at $95\%$~CL. Setting $|D_{21}^{\rm (PV, T)}| \leq \Delta D$, we obtain the exclusion plots in Fig.~\ref{fig_exc_hfs}. The central upward peaks are due to the structure of Sternheim's splitting, which introduces the difference of the hyperfine splittings of the $1s$ and $2s$ states. From Eqs.~\eqref{eq_D_21_PV} and~\eqref{eq_D_21_T} we see that both peaks occur at $m^\star = ( \sqrt{10} - 2 )k/3 \approx 1.45$~keV. For the PV coupling the limit behaves as $1/m^2$ for $m \ll m^\star$ and as $m^2$ for $m \gg m^\star$, whereas for the T coupling the limit is constant for $m \ll m^\star$ and goes as $m^4$ for $m \gg m^\star$.

The worsening of the bounds for large masses is expected, since a heavy mediator can be hardly excited at low energies and the photon-mediated interaction dominates, thus causing the bound to weaken. The case of small masses is more involved. A truly massless theory is fundamentally different than a massive one, even if the mediator is light relative to the pertinent energy scales: the former is gauge invariant and the latter is not. Let us then focus on the case of a light, but massive, mediator. In this regime the PV potential~\eqref{eq_pot_PV_KRLP} reduces to the standard Maxwellian dipole-dipole interaction~\cite{Griffiths_hyp}, apart from an overall constant. This means that, for small KRLP masses, the strength of the dipole-dipole interaction in pure QED is simply re-scaled by $\sim(1 + g^2_{\rm PV})$. The consequence is that KRLP exchange becomes unobservable and the PV bound worsens; a similar situation occurs for hidden photons~\cite{Joerg_HP, Kroff_Malta}.

For the bound on T interactions the argument is different. The saturation for small masses can be explained as in Ref.~\cite{Fadeev2} using the toy-model example of the Z boson from the SM with only axial-vector couplings to fermions. In this case the potential also displays a (longitudinal) contribution with a pre-factor $g^2/m_Z^2$, where $g$ is the universal electroweak coupling constant and $m_Z$ is the Z-boson mass. The mass-generation mechanism involves a Higgs boson with $g$ and $m_Z$ related to its vacuum expectation value $v$ by $g^2/m_Z^2 = 4/v^2$. This shows that $g^2/m^2$ is actually constant for the energy shifts~\eqref{eq_Delta_hfs_T_1s} and~\eqref{eq_Delta_hfs_T_2s} to leading order. However, in Fig.~\ref{fig_exc_hfs} we use Sternheim's splitting interval $D_{21} = 8 \Delta E_{\rm hfs}(2s) - \Delta E_{\rm hfs}(1s)$, which kills this leading-order behavior of the energy shifts -- the remaining expression is constant for small $m$, as visible in the plot.

Finally, it is worth noting the similarity of the shape of our bounds to those obtained in Ref.~\cite{Fadeev2}, specially their Figs.~5 and~10 for Sternheim's interval $D_{21}$ in hydrogen, where the authors study generic spin-dependent potentials between spin-1/2 sources with couplings to pseudoscalar and pseudovector mediators. The reason for this resemblance is the fact that the potentials considered, their $V_{\rm pp}$ and 
$V_{\rm AA}$ -- or rather $\mathcal{V}_3$ -- are functionally similar to our $V_{\rm PV}$ and $V_{\rm T}$, respectively.

\subsection{Tree-level unitarity in $e^- + e^+ \rightarrow \ell^- + \ell^+$ ($\ell \neq e$) scattering }  \label{sec_unitarity}
\indent

In this and the next section we focus on the scatterings $e^- + e^+ \rightarrow \ell^- + \ell^+$ for $\ell = \{ e, \mu, \tau \}$, which are well described by pure QED at energies up to the vicinity of the Z-pole, where electroweak (EW) effects become relevant. Particularly important is Bhabha scattering, $e^- + e^+ \rightarrow e^- + e^+$, which is frequently used as a luminosity monitor to calibrate cross-section measurements at lepton colliders~\cite{LEP, SLD, Likic, Bicer, Baer}. Corrections from diagrams with Z-boson exchange alter the QED differential cross section at the level of $1-2\%$~\cite{Derrick} and generally improve the fit of the theory to the data -- these processes were crucial in the early studies of the properties of the Z boson~\cite{Levi, Abrams1, Abrams2}. The EW corrections are of the same order of magnitude as the experimental errors, so we shall include $\ell\ell Z$ vertices along those of QED in our analysis, where the KRLP-mediated diagrams with either PV or T vertices will enter as small perturbations.

The processes $e^- + e^+ \rightarrow \ell^- + \ell^+$ with $\ell = e$ or $\ell \neq e$ are physically distinct, but have some common aspects. Here we study tree-level unitarity applied to the case $\ell \neq e$ and in Sec.~\ref{sec_bhabha} we analyse Bhabha scattering ($\ell = e$). Below we state the necessary kinematics and the amplitudes for the $t$- and $s$-channels; only the latter is relevant for the discussion of tree-level unitarity in Sec.~\ref{sec_unitarity_limits}, whereas both are needed to describe Bhabha scattering in Sec.~\ref{sec_bhabha}.

\subsubsection{Kinematic definitions and amplitudes} \label{sec_kinematics}
\indent

We work in the center-of-mass (CM) frame, where the particles have the same energy $E \gg m_\ell$ and opposite 3-momenta. The 4-momenta of the leptons are the following: the in-coming electron has $p_1 = (E,\textbf{p})$ and the in-coming positron has $p_2 = (E,-\textbf{p})$, whereas the out-going lepton has $p_3 = (E,\textbf{p}')$ and the out-going anti-lepton has $p_4 = (E,-\textbf{p}')$. The CM energy is $\sqrt{s} = 2E$ and $|{\bf p}| = |{\bf p}'|$. The initial 3-momenta are taken to be on the $z$-axis, $ \textbf{p} = \left( \sqrt{s}/2 \right) \, {\bf \hat{z}}$, with the final momentum $\textbf{p}' = \left( \sqrt{s}/2 \right) \, (\sin\theta \cos\phi, \sin\theta \sin\phi, \cos\theta)$.

\begin{figure}[t!]
\centering
\begin{minipage}[b]{0.6\linewidth}
\includegraphics[width=\textwidth]{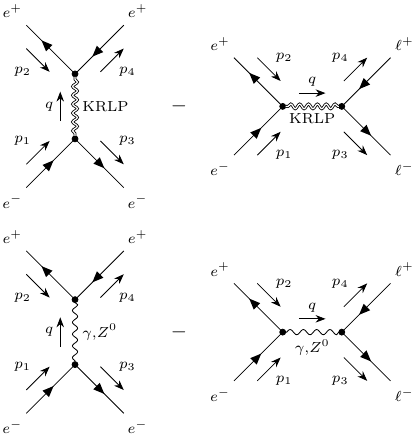}
\end{minipage} \hfill
\caption{ Tree-level Feynman diagrams contributing to the scattering $e^- + e^+ \rightarrow \ell^- + \ell^+$ for $\ell = \{ e, \mu, \tau \}$. If $\ell = e$ we have Bhabha scattering with both $s$- and $t$-diagrams, whereas for $\ell = \{\mu, \tau \}$ only the $s$-channel contributes at tree level. The minus signs indicate that the corresponding amplitudes must be subtracted. }
\label{fig_diagrams}
\end{figure}

The relevant Feynman diagrams are illustrated in Fig.~\ref{fig_diagrams}. The total tree-level amplitudes are the combination of the SM contribution and that of new physics containing KRLP-mediated PV or T interactions: $\mathcal{M}_{\rm tot} = \mathcal{M}_{\rm SM} + \mathcal{M}_{\rm PV, T}$. The amplitude from the SM is itself the sum of the photon-mediated amplitudes
\begin{eqnarray} 
\mathcal{M}_{\rm QED}^{(t)} & = & -\dfrac{e^2}{t} \Big[\bar{u}^{(s_3)}(p_3) \gamma^\mu u^{(s_1)}(p_1) \Big]  \Big[\bar{v}^{(s_2)}(p_2)  \gamma_\mu v^{(s_4)}(p_4) \Big] \, , \label{eq_amp_bhabha_qed_t} \\
\mathcal{M}_{\rm QED}^{(s)} & = & \dfrac{e^2}{s} \Big[\bar{v}^{(s_2)}(p_2) \gamma^\mu u^{(s_1)}(p_1) \Big] \Big[\bar{u}^{(s_3)}(p_3) \gamma_\mu  v^{(s_4)}(p_4) \Big]  \label{eq_amp_bhabha_qed_s}
\end{eqnarray}
with $t = -s\left( 1 - \cos\theta \right)/2$ and of the Z-mediated amplitudes 
\begin{eqnarray} 
\mathcal{M}_{\rm EW}^{(t)} & = & - \dfrac{ g_Z^2}{4(t-m_Z^2 + im_Z \Gamma_Z)}\bigg[\eta_{\mu\nu}  - \dfrac{(p_1-p_3)_\mu (p_1-p_3)_\nu}{m_Z^2} \bigg] \nonumber \\
& \times & \Big[\bar{u}^{(s_3)}(p_3) ( g_V - g_A \gamma_5) \gamma^\mu  u^{(s_1)}(p_1) \Big]  \Big[\bar{v}^{(s_2)}(p_2) ( g_V - g_A \gamma_5) \gamma^\nu v^{(s_4)}(p_4) \Big] \, , \label{eq_amp_bhabha_ew_t} \\
\mathcal{M}_{\rm EW}^{(s)} & = & \dfrac{ g_Z^2}{4(s-m_Z^2 + im_Z \Gamma_Z)}  \bigg[\eta_{\mu\nu}  - \dfrac{(p_1+p_2)_\mu (p_1+p_2)_\nu}{m_Z^2} \bigg] \nonumber \\
& \times & \Big[\bar{v}^{(s_2)}(p_2) ( g_V - g_A \gamma_5) \gamma^\mu u^{(s_1)}(p_1) \Big] \Big[\bar{u}^{(s_3)}(p_3)( g_V - g_A \gamma_5) \gamma^\nu v^{(s_4)}(p_4) \Big] \, . \label{eq_amp_bhabha_ew_s}
\end{eqnarray}

The $\ell\ell Z$ coupling is $g_Z = e/\left( \sin\theta_W \cos\theta_W \right)$, where $\sin^2\theta_W = 0.23$ is the Weinberg angle, $\Gamma_Z = 2.5$~GeV is the Z-boson width and $G_F = 1.17 \times 10^{-5} \, {\rm GeV}^{-2}$ is the Fermi constant~\cite{PDG}. Furthermore, $g_V = -1/2+ 2\sin^2\theta_W$ and $g_A = +1/2$ are the vector and axial couplings.
The longitudinal pieces of the Z-boson propagator are negligible by virtue of the Dirac equation: for the incoming electron we have $p_1^\mu \gamma_\mu u^{(s_1)}(p_1) = m_e u^{(s_1)}(p_1)$. Similar results hold for the other leptons and the longitudinal term contributes with $\sim m_\ell^2/m_Z^2 \lesssim 10^{-4}$, being henceforth ignored; see Refs.~\cite{Chanowitz_1, Chanowitz_2} for a complementary discussion of ultra-heavy fermions.

The KRLP-mediated $t$- and $s$-channel diagrams with PV vertices are
\begin{eqnarray} 
\mathcal{M}_{\rm PV}^{(t)} & = & -\dfrac{ g_{\rm PV}^{e} g_{\rm PV}^{e}  }{2\Lambda^2 (t-m^2)} \bigg\{ t \Big[\bar{u}^{(s_3)}(p_3)  \gamma_5 \gamma_\mu u^{(s_1)}(p_1) \Big]  \Big[\bar{v}^{(s_2)}(p_2) \gamma_5 \gamma^\mu  v^{(s_4)}(p_4) \Big]  \nonumber \\
 & - & q_\mu q_\nu \Big[\bar{u}^{(s_3)}(p_3) \gamma_5 \gamma^\mu u^{(s_1)}(p_1) \Big]  \Big[\bar{v}^{(s_2)}(p_2) \gamma_5 \gamma^\nu  v^{(s_4)}(p_4) \Big] \bigg\}  \, , \label{eq_amp_PV_t} \\
\mathcal{M}_{\rm PV}^{(s)} & = &  \dfrac{ g_{\rm PV}^{e} g_{\rm PV}^{\ell} }{2\Lambda^2 (s-m^2)}  \bigg\{ s \Big[\bar{v}^{(s_2)}(p_2) \gamma_5 \gamma_\mu u^{(s_1)}(p_1) \Big] \Big[\bar{u}^{(s_3)}(p_3) \gamma_5 \gamma^\mu  v^{(s_4)}(p_4) \Big]
\nonumber \\
 & - & q_\mu q_\nu  \Big[\bar{v}^{(s_2)}(p_2) \gamma_5 \gamma^\mu  u^{(s_1)}(p_1) \Big] \Big[\bar{u}^{(s_3)}(p_3) \gamma_5 \gamma^\nu v^{(s_4)}(p_4) \Big] \bigg\} \,  \label{eq_amp_PV_s}
\end{eqnarray}
and the KRLP-mediated amplitudes with T vertices are
\begin{eqnarray} 
\mathcal{M}_{\rm T}^{(t)} & = & - \dfrac{ g_{\rm T}^{e} g_{\rm T}^{e} }{(t-m^2)}\bigg\{   \Big[\bar{u}^{(s_3)}(p_3) \sigma_{\mu\nu} u^{(s_1)}(p_1) \Big]  \Big[\bar{v}^{(s_2)}(p_2) \sigma^{\mu \nu} v^{(s_4)}(p_4) \Big]  \nonumber \\
& + & \frac{2q^\nu q_\alpha}{m^2} \Big[\bar{u}^{(s_3)}(p_3) \sigma_{\mu\nu} u^{(s_1)}(p_1) \Big]  \Big[\bar{v}^{(s_2)}(p_2) \sigma^{\mu \alpha} v^{(s_4)}(p_4) \Big]  \bigg\}  \, , \label{eq_amp_T_t} \\
\mathcal{M}_{\rm T}^{(s)} & = & \dfrac{ g_{\rm T}^{e} g_{\rm T}^{\ell} }{(s-m^2)}\bigg\{   \Big[\bar{v}^{(s_2)}(p_2) \sigma_{\mu\nu} u^{(s_1)}(p_1) \Big]  \Big[\bar{u}^{(s_3)}(p_3) \sigma^{\mu \nu} v^{(s_4)}(p_4) \Big]  \nonumber \\
& + & \frac{2q^\nu q_\alpha}{m^2} \Big[\bar{v}^{(s_2)}(p_2) \sigma_{\mu\nu} u^{(s_1)}(p_1) \Big]  \Big[\bar{u}^{(s_3)}(p_3) \sigma^{\mu \alpha} v^{(s_4)}(p_4) \Big]  \bigg\} \, . \label{eq_amp_T_s}
\end{eqnarray}
If $\ell \neq e$ only the $s$-channel diagrams are allowed and if $\ell = e$ we have Bhabha scattering with the two channels. Here $q$ is the 4-momentum carried by the KRLP: for the $s$-channel $q = p_1 + p_2 = p_3 + p_4$ and $q^2 = s$, while for the $t$-channel $q = p_1 - p_3 = -(p_2 - p_4)$ and $q^2 = t$.

\subsubsection{Perturbative unitarity for $\ell \neq e$} \label{sec_unitarity_limits}
\indent 

The scattering matrix $S = 1 + iT$ describes the interaction of particles. Non-trivial interactions are encoded in the transition matrix $T$, which takes the system from the initial state $|i\rangle$ to the final state $|f\rangle$. Its matrix elements, $T_{if} = \langle i | T | f \rangle$, are expressed in terms of the Lorentz-invariant amplitude $\mathcal{M}_{i \rightarrow f}$ as $T_{if} =\left( 2\pi \right)^4 \delta^4\left( P_i - P_f\right) \mathcal{M}_{i \rightarrow f}$, where $P_{i,f}$ are the sum of 4-momenta of the $|i\rangle, |f\rangle$ states, respectively.

The amplitude for $e^-(h) + e^+(\overline{h}) \rightarrow \ell^-(h') + \ell^+(\overline{h'})$ may be labeled by the helicities of the particles as $\mathcal{M}_{h\overline{h} \,\rightarrow\, h'\overline{h'}}$ with $\{ h,\overline{h},h',\overline{h'}\} = \pm 1$ (or $\pm$ for short); these indicate the eigenvalues of the helicity operator, cf.~App.~\ref{app_helicity_spinors}. For a given helicity configuration, the amplitude depends on the CM energy and on the scattering angle. It is convenient to decompose it in terms of partial waves of total angular momentum $J$ in the limit of high energies via~\cite{Wick, Logan} 
\begin{equation} \label{eq_amp_partial_wave}
\mathcal{M}_{h\overline{h} \,\rightarrow\, h'\overline{h'}}(\sqrt{s}, \cos\theta) = 16\pi \sum_{J = 0}^\infty \left( 2J + 1 \right) d^{J}_{\mu_i,\mu_f}(\theta) a^J_{h\overline{h} \,\rightarrow\, h'\overline{h'}} (\sqrt{s})  \, .
\end{equation}
Here $d^J_{\mu_i,\mu_f} (\theta)$ are the Wigner small-$d$ functions with $\mu_i = \frac{1}{2}(h - \overline{h})$, $\mu_f = \frac{1}{2}(h' - \overline{h'})$ and $\mu_{i,f} = -J, -J + 1, \cdots, J - 1, J$~\cite{Wigner}.

The generally complex coefficients $a^{J}_{h\overline{h} \,\rightarrow\, h'\overline{h'}}(\sqrt{s})$, the partial waves, may be expressed in terms of the amplitude as~\cite{Wick, Luzio}
\begin{equation} \label{eq_partial_wave}
a^{J}_{h\overline{h} \,\rightarrow\, h'\overline{h'}} (\sqrt{s}) = \frac{1}{32\pi} \int_{-1}^1 d(\cos\theta) d^J_{\mu_i,\mu_f} (\theta) \mathcal{M}_{h\overline{h} \,\rightarrow\, h'\overline{h'}}(\sqrt{s}, \cos\theta) \, .
\end{equation}
Limiting the number of particles in the intermediate states to two, the same as in $|i\rangle$ and $|f\rangle$, the real and imaginary parts of the partial wave~\eqref{eq_partial_wave} are bounded from above by~\cite{Wick, Logan} 
\begin{equation} \label{eq_unit_bounds}
{\rm Re} \left( a^{J}_{h\overline{h} \,\rightarrow\, h'\overline{h'}} \right) \leq \frac{1}{2} \quad {\rm and} \quad {\rm Im} \left( a^{J}_{h\overline{h} \,\rightarrow\, h'\overline{h'}} \right) \leq 1 \, .
\end{equation}
These are the unitarity bounds, from which the first inequality is the most useful, since tree-level amplitudes are real. If more than one amplitude contributes to a certain partial wave, the bounds above are applied to the (absolute value of the) largest eigenvalue.

These bounds are satisfied by the complete, renormalizable theory from which the non-perturbative amplitude is calculated. We are, however, pursuing a perturbative treatment, but we may still apply the bounds~\eqref{eq_unit_bounds} to each term in the series expansion of the amplitude. If the tree-level amplitude violates unitarity and we know that a complete theory should respect it, we have an indication that higher-order terms must restore unitarity: the perturbative expansion in the coupling parameter is not reliable, since the lower-order terms are not a good approximation of the full amplitude. Thus, the bounds~\eqref{eq_unit_bounds} represent limits on the perturbative unitarity and may be used to fix the range of validity of the parameters of an effective theory.

A classical example is the ``Higgsless" EW theory. The scattering of longitudinal W bosons would violate unitarity at tree level -- the four-point interaction amplitudes and $s$- and $t$-channel photon and Z-boson exchange grow as $\sim s/v_F^2$, where $v_F \approx 246$~GeV. Therefore, in a ``Higgsless" SM, perturbative unitarity would be violated in high energies~\cite{Logan, Chano_WW}. If, however, a scalar Higgs boson with mass $m_H$ is included, the added amplitudes with Higgs exchange exactly cancel and the aforementioned problem is avoided.

Moreover, taking other channels which proceed via Higgs exchange into account, the final amplitude turns out to be $\sim m_H^2/v_F^2$. Once again, applying Eq.~\eqref{eq_unit_bounds} one finds that $m_H \lesssim 710$~GeV~\cite{SM_unit_1, SM_unit_2}, thus providing an upper limit on the Higgs mass based on perturbative unitarity. Naturally, corrections at higher orders may also break the perturbative expansion~\cite{Bij_Higgs}. A similar argument (at tree level) was responsible for the introduction of intermediate vector bosons in the context of the old Fermi theory of weak interactions~\cite{djouadi}: above the Fermi scale new degrees of freedom -- the massive W and Z bosons -- were expected to emerge if (perturbative) unitarity was to be respected.

Let us move on to the case of interest, $e^- + e^+ \rightarrow \ell^- + \ell^+$ with $\ell \neq e$. Working in the ultra-relativistic regime, where $\sqrt{s} \gg m_\ell$, the leptons are essentially massless and in App.~\ref{app_helicity_spinors} we construct the corresponding spinors, cf. Eq.~\eqref{eq_spinors_u_v_UR}, which are eigenvectors of the helicity operator with eigenvalues $h$. The leptons are propagating with the following angular parameters $\{ \theta, \phi \}$: incoming $e^-(p_1, h)$ with $\{ 0, 0 \}$, incoming $e^+(p_2, \overline{h})$ with $\{ \pi, \pi \}$, outgoing $\ell^-(p_3, h')$ with $\{ \theta, 0 \}$ and outgoing $\ell^+(p_4, \overline{h'})$ with $\{ \pi - \theta, \pi \}$. The amplitudes may thus be naturally labelled $\mathcal{M}_{h\overline{h} \,\rightarrow\, h'\overline{h'}}$. For $\ell \neq e$ only the $s$-channel diagrams contribute, cf. Fig.~\ref{fig_diagrams}, and the non-zero helicity amplitudes for pure QED, cf. Eq.~\eqref{eq_amp_bhabha_qed_s}, are 
\begin{eqnarray}
\mathcal{M}_{+- \,\rightarrow\, +-}^{\rm QED} & = &  \mathcal{M}_{-+ \,\rightarrow\, -+}^{\rm QED} = e^2 (1 + \cos\theta)  \, , \label{eq_hel_amp_qed_1}\\
\mathcal{M}_{+- \,\rightarrow\, -+}^{\rm QED} & = &  \mathcal{M}_{-+ \,\rightarrow\, +-}^{\rm QED} = e^2(1 - \cos\theta) \, . \label{eq_hel_amp_qed_2}
\end{eqnarray}
For the Z-mediated amplitude, cf. Eq.~\eqref{eq_amp_bhabha_ew_s}, neglecting the width of the Z boson and setting $z = \sqrt{s}/m_Z$, we have
\begin{eqnarray}
\mathcal{M}_{+- \,\rightarrow\, +-}^{\rm EW} & = &  \frac{  g_Z^2 \left(g_V + g_A \right)^2  z^2 (1 + \cos\theta) }{ 4(1 - z^2) }  \underset{z \gg 1}{\longrightarrow}  -\frac{  g_Z^2 \left(g_V + g_A \right)^2  (1 + \cos\theta) }{ 4 } \, , \label{eq_hel_amp_ew_1} \\
\mathcal{M}_{-+ \,\rightarrow\, -+}^{\rm EW} & = &  \frac{  g_Z^2 \left(g_V - g_A \right)^2  z^2 (1 + \cos\theta) }{ 4(1 - z^2) }  \underset{z \gg 1}{\longrightarrow}  -\frac{  g_Z^2 \left(g_V - g_A \right)^2  (1 + \cos\theta) }{ 4 }  \, , \label{eq_hel_amp_ew_2} \\
\mathcal{M}_{+- \,\rightarrow\, -+}^{\rm EW} & = & \mathcal{M}_{-+ \,\rightarrow\, +-}^{\rm EW} = \frac{ g_Z^2 (g_V^2 - g_A^2) z^2 \left( 1 - \cos\theta \right)}{ 4\left( 1 - z^2 \right) }  \underset{z \gg 1}{\longrightarrow}  -\frac{ g_Z^2 (g_V^2 - g_A^2) \left( 1 - \cos\theta \right)}{ 4 } \, . \label{eq_hel_amp_ew_3}
\end{eqnarray}

\newpage
The KRLP-mediated amplitudes, Eqs.~\eqref{eq_amp_PV_s} and~\eqref{eq_amp_T_s}, have a more complex momentum structure involving $q_\mu = (p_1 + p_2)_\mu = \sqrt{s} (1, 0, 0, 0)$. With this, the longitudinal term becomes $(p_1 + p_2)_\mu (p_1 + p_2)_\nu = {\rm diag}(s,0,0,0)$ and the PV helicity amplitudes read ($y = \sqrt{s}/m$)
\begin{eqnarray}
\mathcal{M}_{+- \,\rightarrow\, +-}^{\rm PV} & \!\! = \!\! & \mathcal{M}_{-+ \,\rightarrow\, -+}^{\rm PV} = \left( \frac{ g_{\rm PV}^{e} g_{\rm PV}^{\ell} }{\Lambda^2} \right) \frac{  m^2 y^4 (1 + \cos\theta) }{ 2(1 - y^2) }  \underset{y \gg 1}{\longrightarrow}  -\left( \frac{ g_{\rm PV}^{e} g_{\rm PV}^{\ell} }{\Lambda^2} \right) \frac{  s (1 + \cos\theta) }{ 2 }, \label{eq_hel_amp_PV_1} \\
\mathcal{M}_{+- \,\rightarrow\, -+}^{\rm PV} & \!\! = \!\! & \mathcal{M}_{-+ \,\rightarrow\, +-}^{\rm PV} = -\left( \frac{ g_{\rm PV}^{e} g_{\rm PV}^{\ell} }{\Lambda^2} \right) \frac{m^2 y^4 (1 - \cos\theta) }{ 2(1 - y^2) }  \underset{y \gg 1}{\longrightarrow}    \left( \frac{ g_{\rm PV}^{e} g_{\rm PV}^{\ell} }{\Lambda^2} \right) \frac{  s (1 - \cos\theta) }{ 2 }  \label{eq_hel_amp_PV_2}
\end{eqnarray}
and 
\begin{eqnarray}
\mathcal{M}_{++ \,\rightarrow\, ++}^{\rm T} & = & \mathcal{M}_{-- \,\rightarrow\, --}^{\rm T} = \left( g_{\rm T}^{e} g_{\rm T}^{\ell}  \right) \frac{2y^2 (2 + y^2) \cos\theta}{(1 - y^2)}  \underset{y \gg 1}{\longrightarrow}  -\left( g_{\rm T}^{e} g_{\rm T}^{\ell}  \right) \frac{2s\cos\theta}{m^2} , \label{eq_hel_amp_T_1} \\
\mathcal{M}_{++ \,\rightarrow\, --}^{\rm T} & = & \mathcal{M}_{-- \,\rightarrow\, ++}^{\rm T} = \left( g_{\rm T}^{e} g_{\rm T}^{\ell}  \right) \frac{2y^4 \cos\theta}{(1 - y^2)}   \underset{y \gg 1}{\longrightarrow}  -\left( g_{\rm T}^{e} g_{\rm T}^{\ell}  \right) \frac{2s\cos\theta}{m^2}  \,  . \label{eq_hel_amp_T_2}
\end{eqnarray}
All other amplitudes are either identically zero in the $s$-channel or vanish in the ultra-relativistic limit. The QED and EW amplitudes are independent of the CM energy, whereas the KRLP-mediated amplitudes grow with $\sim s$. The physical reason behind this is two-fold. First we have that the fermion currents in QED and EW are conserved; for the PV and T currents this is not so, cf. Eqs.~\eqref{eq_dj_pv} and~\eqref{eq_dj_t}. The second is the much richer momentum structure of the KRLP propagator in comparison with its QED and EW counterparts. These two factors connected give rise to extra 4-momentum contributions in the amplitudes. Furthermore, in the high-energy limit, the KRLP mass is absent from the amplitudes involving PV vertices, but not from those with T vertices.

In the CM frame, the initial and final $\ell \overline{\ell}$ pairs have the same energies, moving with equal and opposite 3-momenta; the system has zero orbital angular momentum. Also, in the ultra-relativistic regime the spins are aligned parallel or anti-parallel to the particle's 3-momentum. Since the external states are composed of pairs of spin-1/2 fermions, we have total angular momentum (projection) $J=0,1$. For $J = 0$ we have $\mu_{i,f} = 0$ and only the helicities $\{h \, \overline{h}\} = \{++,--\}$ are compatible -- the relevant amplitudes are $\mathcal{M}_{++ \,\rightarrow\, ++}$, $\mathcal{M}_{-- \,\rightarrow\, --}$ and $\mathcal{M}_{++ \,\leftrightarrow\, --}$. The allowed helicity combinations can be found only in the KRLP-mediated T amplitudes, cf. Eqs.~\eqref{eq_hel_amp_T_1} and~\eqref{eq_hel_amp_T_2}, since the corresponding QED, EW and PV amplitudes are zero. However, using Eq.~\eqref{eq_partial_wave} with $J = 0$, for which $d^0_{0,0}(\theta) = 1$, we find that $a^{J=0}_{\rm T} = 0$. Therefore, no useful bounds on the PV and T couplings can be derived from perturbative unitarity in the sector $J = 0$.

The sector $J=1$ is much richer, because now $\mu_{i,f} = 0,\pm 1$. Let us start by the KRLP-mediated PV interaction. We already know that the sub-sector $\mu_{i,f} = 0$ does not contribute to the partial wave, but the sub-sectors $\mu_{i,f} = \pm 1$ do -- only amplitudes with initial and final states with helicities $\{ h \overline{h} \} = \{ +-, -+ \}$ are relevant in this case. From Eq.~\eqref{eq_partial_wave} with $J = 1$ and using that $d^1_{1,1}(\theta) = d^1_{-1,-1}(\theta) = (1 + \cos\theta)/2$ and $d^1_{1,-1}(\theta) = d^1_{-1,1}(-\theta) = (1 - \cos\theta)/2$ the partial-wave matrix for the PV coupling, in the basis $\{ +-, -+ \}$, becomes
\begin{equation} 
a^{J=1}_{\rm PV} \approx \frac{s}{48\pi} \left( \frac{ g_{\rm PV}^{e} g_{\rm PV}^{\ell} }{\Lambda^2} \right) \left( \begin{array}{cc}
-1   & 1 \\
1   &  -1
 \end{array} \right) \, . \label{eq_coupled_channel_PV_1} 
\end{equation}
Applying Eq.~\eqref{eq_unit_bounds} to the absolute value of the largest eigenvalue of $a^{J=1}_{\rm PV}$, we get
\begin{equation} \label{eq_unitarity_bound_PV}
\left| \frac{ g_{\rm PV}^{e} g_{\rm PV}^{\ell} }{\Lambda^2} \right| \leq \frac{12\pi}{s} = 3.8 \times 10^{-17} \, {\rm eV}^{-2} \left( \frac{{\rm GeV}}{\sqrt{s}} \right)^2 \, .
\end{equation}
Since we assumed very high energies, in particular $\sqrt{s} \gg m_Z$, the result above can be more simply stated as $| g_{\rm PV}^{e} g_{\rm PV}^{\ell}/\Lambda^2 | \ll 4.6 \times 10^{-21} \, {\rm eV}^{-2}$. The PV coupling constant $g_{\rm PV}^f$ is dimensionless and we may as well assume it to be $\sim \mathcal{O}(1)$, that is, we allow $\Lambda$ to determine the overall magnitude (and canonical dimension) of the PV coupling $g_{\rm PV}^f/\Lambda$. In this context the bound~\eqref{eq_unitarity_bound_PV} may be converted into $\Lambda \geq 1.6 \times 10^{8} \, {\rm eV} \left( \frac{\sqrt{s}}{{\rm GeV}} \right)$. 

%

Finally, let us analyse the KRLP-mediated interactions involving T couplings that contribute to the sector $J = 1$. As already mentioned, only the sub-sector $\mu_{i,f} = 0$ with helicities $\{h \, \overline{h}\} = \{++,--\}$ is relevant and the QED and EW amplitudes are zero. Using Eqs.~\eqref{eq_hel_amp_T_1} and~\eqref{eq_hel_amp_T_2} together with $d^1_{0,0}(\theta) = \cos\theta$, we find that the partial-wave matrix in the basis $\{++,--\}$ is
\begin{eqnarray} 
a^{J=1}_{\rm T} & \approx & -\frac{s}{24 \pi} \left( \frac{ g_{\rm T}^{e} g_{\rm T}^{\ell} }{m^2} \right) \left( \begin{array}{cc}
1   & 1 \\
1   &  1
 \end{array} \right) \, ,\label{eq_coupled_channel_PV_1} 
\end{eqnarray}
so that, taking the absolute value of the largest eigenvalue and imposing~\eqref{eq_unit_bounds}, we get
\begin{equation} \label{eq_unitarity_bound_T}
|g_{\rm T}^{e} g_{\rm T}^{\ell}|  \leq 6\pi \left( \frac{m}{\sqrt{s}} \right)^2 = 1.9 \times 10^{-17} \left( \frac{m}{{\rm eV}} \right)^2 \left( \frac{{\rm GeV}}{\sqrt{s}} \right)^2 \, .
\end{equation}
Once more, the result above was obtained under the condition that $\sqrt{s} \gg m_Z$, and it can thus be translated into $|g_{\rm T}^{e} g_{\rm T}^{\ell}|  \ll 2.3 \times 10^{-21} \left( m/{\rm eV} \right)^2$.

Contrary to the PV case, here we are not allowed to assume $g_{\rm T}^f \sim \mathcal{O}(1)$ in order to convert the limit above into a more direct limit on $m$. The reason is that $g_{\rm T}^f$ is the actual (dimensionless) coupling of the T interaction, cf. Eq.~\eqref{eq_ints}, that is used to establish the perturbative expansion that is ultimately constrained by unitarity. The PV interaction is parametrized not only by $g_{\rm PV}^f$, but by the (energy) parameter $\Lambda$ in the combination $g_{\rm PV}^f/\Lambda$, and in this case we may assume that $g_{\rm PV}^f \sim \mathcal{O}(1)$ and transfer the size of the PV interaction to $\Lambda$. This is not so for $g_{\rm T}^f$ and the KRLP mass $m$ cannot be bounded independently.

\subsection{Bhabha scattering}  \label{sec_bhabha}
\indent

In Sec.~\ref{sec_kinematics} we stated the amplitudes and obtained limits on the PV and T couplings from perturbative unitarity. Now we move on to study the unpolarized differential cross section of Bhabha scattering mediated by KRLPs. Before stating the final result including the new interactions, let us first quote the expected background result from the SM, which includes photon- and Z-mediated tree-level diagrams as shown in Fig.~\ref{fig_diagrams}.

Working in the ultra-relativistic limit ($\sqrt{s} \gg m_e$), the unpolarized differential cross section from EW theory is~\cite{Derrick} 
\begin{equation} \label{eq_dcs_tot_SM}
\frac{d \sigma_{\rm SM}}{d(\cos\theta)} = \frac{\pi\alpha^2}{s} \left[ |A(t)|^2 \left( \frac{s}{t} \right)^2 + |A(s)|^2 \left( \frac{t}{s} \right)^2 + \frac{1}{2} \left( |A_{+}|^2 + |A_{-}|^2 \right) \left( 1 + \frac{t}{s} \right)^2 \right] \, ,
\end{equation}
where $\alpha = e^2/4\pi \approx 1/137$ is the fine structure constant and 
\begin{equation}
A(a) = 1 + \left( g_V^2 - g_A^2 \right)\xi (a) \quad {\rm and} \quad A_\pm = 1 + \frac{s}{t} + \left( g_V \pm g_A \right)^2 \left[ \xi (s) + \frac{s}{t} \xi (t) \right] 
\end{equation}
with $\xi (a) = \frac{G_F}{\pi\alpha\sqrt{8}} \frac{a m_Z^2}{\left( a - m_Z^2 + im_Z \Gamma_Z\right)}$. Here $g_V = -1/2+ 2\sin^2\theta_W$ and $g_A = +1/2$ are the vector and axial couplings of the Z boson to fermions. The QED contribution is obtained from Eq.~\eqref{eq_dcs_tot_SM} turning off  purely EW effects, that is, by setting $g_V = g_A = 0$:
\begin{equation}\label{eq_bhabha_qed} 
\frac{d \sigma_{\rm QED}}{d(\cos\theta)} = \frac{\pi \alpha^2 (3 + \cos^2\theta )^2}{2 s ( 1 - \cos\theta )^2} \, .
\end{equation}

The total differential cross section is
\begin{equation} \label{eq_dcs_0}
\frac{d \sigma}{d(\cos\theta)} = \frac{d \sigma_{\rm SM}}{d(\cos\theta)} + \frac{d \sigma_{\gamma}^{\rm PV, T}}{d(\cos\theta)} + \frac{d \sigma_{\rm Z}^{\rm PV, T}}{d(\cos\theta)} + \frac{d \sigma^{\rm PV, T}}{d(\cos\theta)}  \, .
\end{equation}
The first term is the SM result, Eq.~\eqref{eq_dcs_tot_SM}. The second and third terms represent contributions from the interference of diagrams with KRLP exchange and diagrams with photon and Z-boson exchange, respectively. The fourth term is purely due to KRLP-mediated diagrams. The aforementioned unpolarized differential cross sections are given in App.~\ref{app_dcs} and their general behaviors are summarized in Table~\ref{table_dcs_behavior}.

Due to the non-conservation of the PV and T currents the longitudinal part of the KRLP propagator does not cancel, leading to extra energy factors in the amplitudes causing the (differential) cross sections to generally grow with the CM energy, cf. Table~\ref{table_dcs_behavior}. This is in stark contrast with QED, where $d\sigma/d(\cos\theta) \sim 1/s$, cf. Eq.~\eqref{eq_bhabha_qed}, thus creating an unexpected -- and unobserved -- excess in the number of events at higher energies. This would, in turn, convert into potentially strong bounds on the couplings of KRLPs to electrons and positrons.

\begin{table}
\centering
\begin{tabular}{|c|c|c|c|c|c|c|}
\hline
 & \, QED interf. PV \, &  \, pure PV   \,  &   \, QED interf. T  \, &  \, pure T  \,   \\ \hline\hline
 \, $m \ll \sqrt{s}$  \, & $\sim (g^2/\Lambda^2) \alpha$  & $\sim (g^4/\Lambda^4) s$  &  $\sim g^2 \alpha/m^2$  & $\sim g^4 s/m^4$  \\ \hline
 \, $m \gg \sqrt{s}$  \, & $\sim (g^2/\Lambda^2)\alpha s/m^2$  & $\sim (g^4/\Lambda^4) s^3/m^4$  & $\sim g^2 \alpha s/m^4$  & $\sim g^4 s/m^4$  \\ \hline
\end{tabular}
\caption{ Behavior of the differential cross sections for PV and T couplings as a function of $m$ and $\sqrt{s}$ for the limiting cases of $m \ll \sqrt{s}$ and $m \gg \sqrt{s}$. Angular factors are omitted and the couplings are stripped of super- and subscripts for notational clarity, being denoted simply by $g$. The interference terms with the Z-mediated diagrams follow the same patterns as the interference with QED for both PV and T couplings. }\label{table_dcs_behavior}
\end{table}

\subsubsection{Limits from deviations from pure QED} \label{sec_bhabha_limits_qed}
\indent

Particularly interesting for our purposes is the analysis from Ref.~\cite{Derrick} reporting on dedicated Bhabha measurements at $\sqrt{s} = 29$~GeV in the angular range $|\cos\theta| < 0.55$ performed with the high resolution spectrometer at the PEP $e^+ \, e^-$ storage ring facility at SLAC. One of their main results is the determination of upper bounds on deviations of the data from the predictions of QED (at $95\%$~CL):
\begin{equation}\label{eq_bhabha_cut_1}
\Bigg| \frac{d {\sigma}/d (\cos\theta)}{d \sigma_{\rm QED}/d (\cos\theta)} - 1 \Bigg| \lesssim \frac{3 s}{E_{\rm cut-off}^2} \, 
\end{equation}
with $E_{\rm cut-off} \approx 200$~GeV. This relation may be used to constrain new physics (see e.g. Refs.~\cite{Pedro_LSV, Dutta, Bufalo, Bufalo2}), since the left-hand side is the ratio of the differential cross sections due to new physics relative to that of QED, cf. Eq.~\eqref{eq_bhabha_qed}. We thus have
\begin{equation}\label{eq_bhabha_cut_2}
\Bigg| \frac{d \sigma^{\rm PV, T}_\gamma/d (\cos\theta) + d \sigma^{\rm PV, T}/d (\cos\theta)}{d \sigma_{\rm QED}/d (\cos\theta)} \Bigg| \lesssim 0.063  \, .
\end{equation}

\begin{figure}[t!]
\begin{minipage}[b]{1.0\linewidth}
\includegraphics[width=\textwidth]{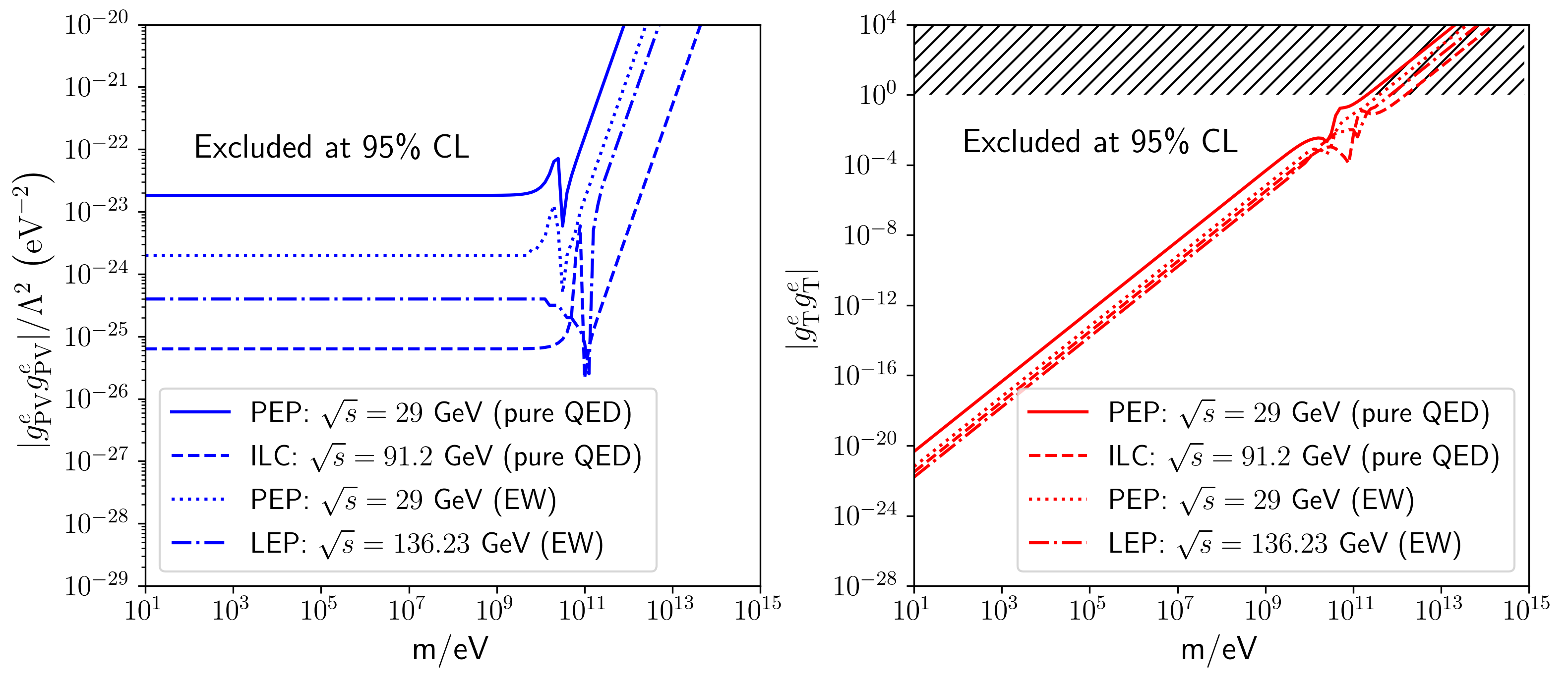}
\end{minipage} \hfill
\caption{ Constraints on PV and T couplings of KRLPs from Bhabha scattering. The bounds obtained by using upper limits for deviations from pure QED are from PEP and the ILC (projected), both labelled ``pure QED", cf. Sec.~\ref{sec_bhabha_limits_qed}. For the data from PEP we used $\sqrt{s} = 29$~GeV~\cite{Derrick}, whereas for the projected sensitivities for the ILC we used $\sqrt{s} = m_Z$~\cite{ILC}. For both we set $\cos\theta = -0.3$. The remaining bounds were extracted using experimental data at various angles from PEP and LEP, cf. Sec.~\ref{sec_bhabha_EW}, at $\sqrt{s} = 29$~GeV~\cite{Derrick} and $\sqrt{s} = 136.23$~GeV~\cite{OPAL}, respectively. All bounds are at $95\%$~CL. The hatched region in the right panel is excluded since there $|g_{\rm T}^{e} g_{\rm T}^{e}| \gtrsim \mathcal{O}(1)$ and the perturbative treatment becomes meaningless.}
\label{fig_bounds_Bhabha}
\end{figure}

Typically, one would neglect the pure PV or T contribution in comparison with the respective interference term with photon-mediated diagrams based on the powers of the coupling constant. However, the differential cross sections are relatively complex functions of the KRLP mass and there may be values of the KRLP couplings and mass for a given energy that violate our intuition. We thus choose to keep the pure PV or T contributions and numerically evaluate the constraint~\eqref{eq_bhabha_cut_2}. The $95\%$-CL upper bounds on the parameters of the KRLP-mediated interactions are shown in Fig.~\ref{fig_bounds_Bhabha}, labelled ``pure QED". For the bounds we use $\cos\theta = -0.3$, maximizing the bound for PV couplings; the results for T couplings are rather insensitive to the angle chosen.

A few features displayed by the bounds in Fig.~\ref{fig_bounds_Bhabha} are noteworthy. First is the change in slope of the bound for the PV couplings, as could be anticipated from Table~\ref{table_dcs_behavior}. In the region $m \lesssim \sqrt{s}$ the PV differential cross sections are independent of $m$, leading to the plateau dominated by the QED interference. Moving towards larger masses, the pure PV term is suppressed and the bound relaxes after passing by $m = \sqrt{s}$, since now Eq.~\eqref{eq_dcs_PV_qed} behaves as $\sim y^2 = s/m^2$; the bound becomes $|g_{\rm PV}^{e} g_{\rm PV}^{e}/\Lambda^2| \sim m^2$.

For the T couplings the situation is different. For $m \lesssim \sqrt{s}$ the pure-T differential cross section is enhanced by $s/m^4$, thus compensating the higher order in the coupling constants relative to the QED interference term -- the bound then goes as $|g_{\rm T}^{e} g_{\rm T}^{e}| \sim m^2$. Around $m \approx \sqrt{s}$ the bound reaches $|g_{\rm T}^{e} g_{\rm T}^{e}| \sim \mathcal{O}(1)$ and, going further, the high-mass region lies in the range $|g_{\rm T}^{e} g_{\rm T}^{e}| \gtrsim \mathcal{O}(1)$. This is problematic: the theory becomes strongly coupled and a perturbative treatment as the one undertaken here is no longer justified. Therefore, the bounds above are not reliable for $|g_{\rm T}^{e} g_{\rm T}^{e}| \gtrsim \mathcal{O}(1)$ -- this is indicated in Fig.~\ref{fig_bounds_Bhabha} by the hatched region.

Future lepton colliders, such as the International Linear Collider (ILC)~\cite{ILC} and other ``Higgs factories"~\cite{Bechtle}, will operate at much higher energies with improved precision. Also at the ILC Bhabha scattering at small angles -- as low as $|\cos\theta | = 0.996$ -- will play a central role in the calibration of cross-section measurements. The ILC is planned to reach $\sqrt{s} = 500$~GeV, but luminosity measurements will be particularly critical during its Giga-Z phase at $\sqrt{s} = m_Z$, where an enhancement in the number of Bhabha events is expected. The target precision in the determination of the number of such events, and consequently also of the (differential) cross section, is $0.1\%$. Setting twice this value as the upper bound on deviations from pure QED to reach $95\%$~CL we obtain the projected bounds shown in Fig.~\ref{fig_bounds_Bhabha}.

\subsubsection{Limits from electroweak data at $\sqrt{s} = 29$~GeV and $136.23$~GeV} \label{sec_bhabha_EW}
\indent

A complementary possibility to constrain KRLP-mediated interactions with Bhabha scattering at the GeV scale is to go beyond QED and analyse the full tree-level EW contributions. The SM result~\eqref{eq_dcs_tot_SM} fits well all data so far and deviations caused by the KRLP-mediated interactions will appear as small corrections that must generally fit within experimental error.

In order to extract bounds on the PV and T couplings and the KRLP mass we will perform simple chi-squared analyses of differential cross section data available from lepton colliders. We scan the parameter space and search for the minimum of
\begin{equation} \label{eq_chi2}
\chi^2 = \sum_{i=1}^{N} \frac{ \left[ y_{\rm obs}^i - y_{\rm th}^i \right]^2 }{ \delta y^2} \, ,
\end{equation}
where $y$ stands for $d\sigma/d\cos\theta$, ``obs" and ``th" label the observed quantities and the theoretical values obtained using Eq.~\eqref{eq_dcs_0}, respectively, $N$ is the number of data points and $\delta y_i$ are the associated experimental errors. We use the data sets available in Table~XII of Ref.~\cite{Derrick} at $\sqrt{s} = 29$~GeV for $| \cos\theta | \leq 0.525$ with $N = 22$ and Table~2 of Ref.~\cite{OPAL} at $\sqrt{s} = 136.23$~GeV for $| \cos\theta | \leq 0.9$ with $N = 9$. In Ref.~\cite{Derrick} the polar angles are quoted as the central value of the respective bins; in contrast, in Ref.~\cite{OPAL} the bins are given and we have used the respective central values. In evaluating Eq.~\eqref{eq_chi2} we assume the SM parameters $\{\alpha, m_Z$, $\Gamma_Z$, $\sin^2\theta_W \}$ to be fixed and given by their latest accepted values~\cite{PDG}.

Following Table~40.2 of Ref.~\cite{PDG}, we set $95\%$-CL limits on $\{ g_{\rm PV}^{e} g_{\rm PV}^{e}/\Lambda^2; m \}$ and $\{ g_{\rm T}^{e} g_{\rm T}^{e}; m \}$ by identifying the parameters that satisfy $\chi^2 - \chi^2_{\rm min} = 5.99$. We find: $\chi^2_{\rm min}/N_{\rm dof} = 1.12$ for PV and $\chi^2_{\rm min}/N_{\rm dof} = 1.14$ for T at $\sqrt{s} = 29$~GeV~\cite{Derrick}, and $\chi^2_{\rm min}/N_{\rm dof} = 3.74$ for PV and $\chi^2_{\rm min}/N_{\rm dof} = 4.13$ fot T at $\sqrt{s} = 136.23$~GeV~\cite{OPAL}, where $N_{\rm dof} = N - 2$. The ensuing $95\%$-CL bounds are shown in Fig.~\ref{fig_bounds_Bhabha}, labeled ``EW", together with those from deviations of pure QED, cf. Sec.~\ref{sec_bhabha_limits_qed}, where visible improvements over the latter are achieved.


\section{Conclusion}  \label{sec_conclusions}
\indent

In this work we analysed the phenomenological impact of a neutral and massive spin-1 boson expressed as an antisymmetric rank-2 tensor coupled to pseudovector and tensor fermionic currents. Since these fundamentally differ from the usual vector current from electromagnetism or the V-A current from the electroweak interactions, we can expect that KRLPs would induce distictive signals that could be experimentally probed. Furthermore, despite the dualities relating it to ALPs and HPs, particular in the massless case, a massive and interacting KRLP is physically distinct due to its parity-transformation properties, as well as the fact that the couplings of the fundamental field and its field-strength tensor to fermions is different than those of ALPs or HPs. These features plainly justify the investigation of KRLPs and their phenomenology.

\begin{figure}[t!]
\begin{minipage}[b]{1.0\linewidth}
\includegraphics[width=\textwidth]{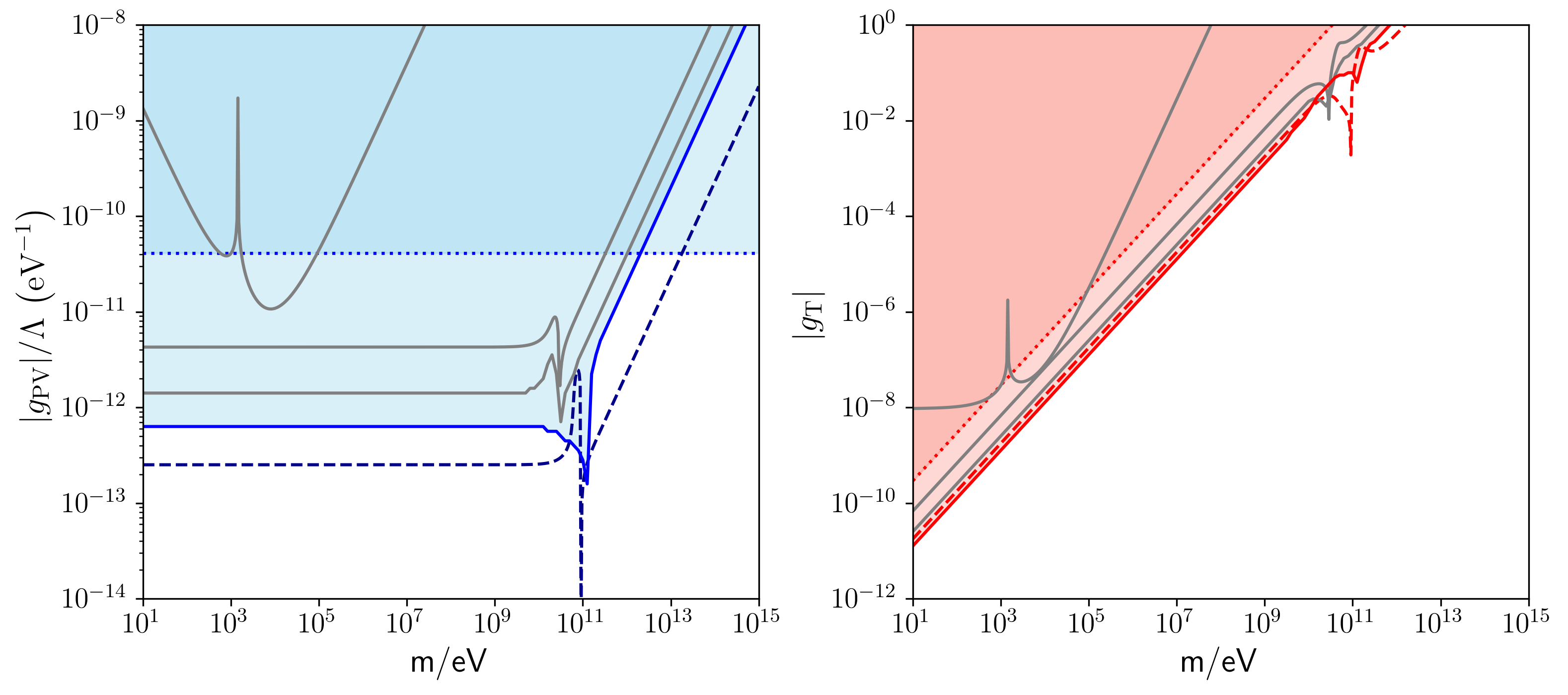}
\end{minipage} \hfill
\caption{ Excluded regions in parameter space for KRLP-mediated interactions assuming that the couplings are independent of the fermion species. The solid lines and the colored regions correspond to the strongest bounds (at $95\%$ CL) coming from Bhabha scattering at $\sqrt{s} = 136.23$~GeV, cf. Sec.~\ref{sec_bhabha_EW}. The projections for Bhabha scattering at the ILC are shown as the dashed lines, cf. Sec.~\ref{sec_bhabha_limits_qed}. The bounds from perturbative unitarity, cf. Sec.~\ref{sec_unitarity_limits}, are shown as dotted lines; the limit from spectroscopy is also shown in grey for completeness. Above $|g_{\rm T}| \sim \mathcal{O}(1)$ the bounds on the right panel are  unreliable. }
\label{fig_global_bounds}
\end{figure}

We studied a variety of systems and processes including KRLPs as small perturbations and analysed how well--measured observables would be affected. In Fig.~\ref{fig_global_bounds} we display the bounds -- all at $95\%$ CL, except those from unitarity -- in case the couplings of the KRLPs to fermions are universal, that is, they are independent of the fermion species: $g^f_{\rm PV, T} = g_{\rm PV, T}$. In general, we found that the scattering amplitudes including KRLPs contain extra factors of the 4-momenta, implying that these amplitudes will be enhanced at high energies. This is visible in Fig.~\ref{fig_global_bounds}, where the bounds from spectroscopy at energy scales $\sim 1/a_0 \sim \mathcal{O}({\rm keV})$ are, despite the outstanding precision in the associated measurements, weaker than those from collider experiments at $\sqrt{s} \sim \mathcal{O}({\rm GeV})$. In fact, the strongest bounds are obtained from LEP data at $\sqrt{s} = 136.23$~GeV~\cite{OPAL}, displayed as solid blue (PV) and red (T) lines.

The next generation of lepton colliders may further improve our bounds, since we observed that the differential cross sections generally increase with the CM energy. Incidentally, this feature is already manifest in the bounds from perturbative unitarity, Eqs.~\eqref{eq_unitarity_bound_PV} and~\eqref{eq_unitarity_bound_T}, showing how the allowed range for the couplings must decrease with $s$ to safeguard unitarity. In Sec.~\ref{sec_bhabha} we also obtained the projected bounds from the ILC at $\sqrt{s} = m_Z$, shown in dashed lines in Figs.~\ref{fig_bounds_Bhabha} and~\ref{fig_global_bounds}. These are stronger than those from LEP due to the improved detection precision of Bhabha events.

There are several proposals for $e^- e^+$ colliders besides the ILC: FCC-ee~\cite{FCC}, CEPC~\cite{CEPC}, CCC~\cite{CCC} and CLIC~\cite{CLIC}; also for $\mu^- \mu^+$ colliders~\cite{muon_c}. These are primarily designed as Z-boson and Higgs factories for precision tests of EW physics. Making the simplifying assumption of coupling universality and $m \ll \sqrt{s}$, from Eqs.~\eqref{eq_bhabha_qed} and~\eqref{eq_bhabha_cut_1} and using Eqs.~\eqref{eq_dcs_PV_qed} and~\eqref{eq_dcs_T}, we see that
\begin{eqnarray}
\frac{g_{\rm PV}}{\Lambda} & \lesssim & 4 \times 10^{-11} \, {\rm eV}^{-1} \left( \frac{\delta}{0.1\%} \right)^{1/2} \left( \frac{ {\rm GeV} }{\sqrt{s}} \right) \, , \\
g_{\rm T} & \lesssim & 9 \times 10^{-11} \left( \frac{ m }{{\rm eV}} \right) \left( \frac{\delta}{0.1\%} \right)^{1/4}  \left( \frac{ {\rm GeV} }{\sqrt{s}} \right) \, ,
\end{eqnarray}
where $\delta$ is the expected precision in the determination of Bhabha events -- these are rough numerical values not taking into account angular factors and EW corrections. Clearly, smaller $\delta$ only mildly improves the bounds and higher energies offer optimal leverage and we expect that, if these proposals come to fruition, stronger limits will be derived.

As a closing remark, let us briefly discuss other portals of KRLPs to SM degrees of freedom. Here we assigned a Proca-like mass to the KRLP, cf. Eq.~\eqref{eq_S_KR}, but this is not the only option. Another possibility is a massless KRLP: in this case the solution to the equations of motion $\partial_\mu H^{\mu\nu\lambda} = 0$ is $H^{\mu\nu\lambda} = \epsilon^{\mu\nu\lambda\kappa}\partial_\kappa a$, where $a$ is a pseudoscalar field, akin to the string-motivated axion~\cite{Witten}. The fundamental field of the theory is now $a$, not $B_{\mu\nu}$ and the kinetic term in the action~\eqref{eq_S_KR} turns into the usual one for a spin-0 field and the interactions with fermions, for example through the PV current, cf. Eq.~\eqref{eq_ints}, involve the dual field-strength tensor $\tilde{H}_\mu \sim \partial_\mu a$, similar to the case of massless axions coupled to electrons~\cite{Liu}. The associated phenomenology will therefore differ from the one analysed here.

Alternatively, we may keep a massive KRLP, but model the mass term differently: instead of the Proca term that does not allow for gauge invariance, we introduce a Chern-Simons-like term~\cite{CS} $\sim m \epsilon^{\mu \nu \alpha \beta} \, A_\mu \partial_\nu B_{\alpha \beta}$, where $B_{\alpha \beta}$ is the KR field and $A_\mu$ is an Abelian field. This topological term is gauge invariant and the pair $\{ A_\mu, B_{\alpha \beta} \}$ carries three degrees of freedom, describing a massive spin-1 field. This scenario has been studied in different contexts: spin-dependent interactions~\cite{Grupo1}, cosmic strings~\cite{CSKR_SUSY}, dark electrodynamics~\cite{Denis}, dualities between Stueckelberg and BF gauge theories~\cite{Smailagic,BF}  and scenarios with Lorentz-symmetry violation~\cite{Nascimento:2014owa}. In this topological formulation the KR field is coupled to a 4-potential, usually associated with the photon, but this is not an absolute requirement: for example, in Ref.~\cite{EW_KR} the authors study dark fermions with a portal to the $U_{\rm Y}(1)$ sector of the SM via dipole couplings (see also Ref.~\cite{Cline}). If we use the weak hypercharge field $Y_\mu$ instead of the electromagnetic 4-potential $A_\mu$, the topological term becomes $m \epsilon^{\mu \nu \alpha \beta} \, Y_\mu \partial_\nu B_{\alpha \beta}$ retaining the EW gauge invariance and, through the mixing of mass eigenstates, resulting in $m \epsilon^{\mu \nu \alpha \beta} \, \left( \cos\theta_W A_\mu -  \sin\theta_W Z_\mu \right) \partial_\nu B_{\alpha \beta}$. We thus generate a direct coupling not only to the photon, but also to the Z boson. The theoretical and phenomenological aspects of this topic will be addressed elsewhere.



\section*{Acknowledgments}
We are grateful to the anonymous referee for constructive comments, to J. A. Helay\"el-Neto for motivating this work and for discussions concerning the KR field, as well as to A. K. Kohara and P. de Fabritiis for helpful comments. P.C.M. and C.A.D.Z. are indebted to Marina and Karoline Selbach, and Lalucha Parizek and Crispim Augusto, respectively, for insightful discussions. This work was financed in part by the Coordena\c{c}\~{a}o de Aperfei\c{c}oamento de Pessoal de N\'ivel Superior - Brasil (CAPES) - Finance Code 001. C.A.D.Z. is partially supported by Conselho Nacional de Desenvolvimento Cient\'ifico e Tecnol\'ogico (CNPq) under the grant no. 310703/2021-2. J. P. S. M. and C. A. D. Z. are also funded by Funda\c{c}\~{a}o Carlos Chagas Filho de Amparo \`a Pesquisa do Estado do Rio de Janeiro (Faperj) under Grants no.  E-03/203.638/2024  (Bolsa de Doutorado Nota 10) and E-26/201{.}447/2021 (Programa Jovem Cientista do Nosso Estado), respectively.


\appendix

\section{Spin projectors and the KRLP propagator} \label{app_A}
\indent

In Sec.~\ref{sec_theory} we obtained the wave operator $\mathcal{O}^{\nu\lambda,\alpha\beta}(k)$, cf. Eq.~\eqref{eq_prop_1}, and mentioned that its inverse is the propagator. The inversion of the operator is best performed by using the spin projectors acting on antisymmetric 2-forms~\cite{CSKR_SUSY, proj_proca_1, proj_proca_2}
\begin{equation} 
\left( P^1_b \right)_{\mu \nu  , \, \rho \sigma } \equiv \frac{1}{2} 
\left( 
\theta_{\mu \rho} \, \theta_{\nu \sigma} - \theta_{\mu \sigma} \, \theta_{\nu \rho}
\right) \, , \label{proj_B_1} 
\end{equation}
\begin{equation} 
\left( P^1_e \right)_{\mu \nu  , \, \rho \sigma } \equiv \frac{1}{2} 
\left( \theta_{\mu \rho} \, \omega_{\nu \sigma} + \theta_{\nu \sigma} \, \omega_{\mu \rho} -
\theta_{\mu \sigma} \, \omega_{\nu \rho} - \theta_{\nu \rho} \, \omega_{\mu \sigma}
\right) \, , \label{proj_B_2} 
\end{equation}
with $\theta_{\mu\nu} = \eta_{\mu \nu} - \frac{k_\mu k_\nu}{k^2}$ and $\omega_{\mu\nu} = \frac{k_\mu k_\nu}{k^2}$. The antisymmetric identity is defined as
\begin{equation}  \label{eq_ident}
\left( P^1_b + P^1_e \right)_{\mu \nu  , \, \rho \sigma } = \frac{1}{2} 
\left(\eta_{\mu \rho} \eta_{\nu \sigma} - \eta_{\mu \sigma} \eta_{\nu \rho} 
\right) \equiv \left( 1^{a.s.} \right)_{\mu\nu,\rho\sigma} \, \end{equation}
and the algebra fulfilled by these operators is:
\begin{equation} 
\left( P^1_b \right)_{\mu \nu  , \, \alpha \beta} 
\left( P^1_b \right)^{\alpha \beta}_{  \, \; \; \, , \, \rho \sigma} = 
\left( P^1_b \right)_{\mu \nu  , \, \rho \sigma} \, , \end{equation} 
\begin{equation} 
\left( P^1_e \right)_{\mu \nu  , \, \alpha \beta} 
\left( P^1_e \right)^{\alpha \beta}_{  \, \; \; \, , \, \rho \sigma} = 
\left( P^1_e \right)_{\mu \nu  , \, \rho \sigma} \, , \end{equation} 
\begin{equation} 
\left( P^1_b \right)_{\mu \nu  , \, \alpha \beta} 
\left( P^1_e \right)^{\alpha \beta}_{  \, \; \; \, , \, \rho \sigma} = 0 \, ,
\end{equation}
\begin{equation} 
\left( P^1_e \right)_{\mu \nu  , \, \alpha \beta} 
\left( P^1_b \right)^{\alpha \beta}_{  \, \; \; \, , \, \rho \sigma} = 0 \, .
\end{equation}
The operator $\mathcal{O}$ may be written in terms of the projectors as (omitting tensor indices for clarity)
\begin{equation}
\mathcal{O} = \left(k^2 - m^2\right) P^1_b -m^2 P^1_e \, .
\end{equation}
The propagator is such that $-i\Delta = \mathcal{O}^{-1} = aP^1_e + bP^1_b$ with $a,b$ free coefficients to be determined. Imposing that $\Delta\mathcal{O} = 1^{a.s.}$ we obtain $a = -1/m^2$ and $b = 1/(k^2 - m^2)$, so that
\begin{equation} \label{eq_prop_app}
\Delta = \frac{i}{k^2 - m^2}\left( P^1_b \right) -  \frac{i}{m^2}\left( P^1_e \right) \, .
\end{equation}

\section{Calculation of the non-relativistic amplitudes} \label{app_B}
\indent

In this appendix we detail the calculation of the NR amplitudes for the KRLP-mediated PV and T interactions. For the sake of generality, we make no prior assumption about the hierarchy of the fermion masses; adequate approximations may be made in a case-by-case basis. The two fermions are labelled $f =a,b$ with initial and final 4-momenta $p_1$ and $p_3$, and $p_2$ and $p_4$, respectively. As stated in the main text, we will express our results as expansions in $|{\bf p}_n|/m_f$ ($n = 1,2,3,4$) and we will keep terms only up to and including second order.

It is convenient to consider the problem in the CM frame, where the total initial and final 3-momenta are zero: ${\bf p}_1 + {\bf p}_2 = {\bf 0} = {\bf p}_3 + {\bf p}_4$. Here we use the average 4-momentum of fermion~$a$, $P = (p_1 + p_3)/2$, and the momentum transfer, $q = p_3 - p_1$, as main variables. In the CM frame, the 3-momenta of the fermion $a$ and $b$ translate to ${\bf p}_1 = {\bf P} - {\bf q}/2$ and ${\bf p}_3 = {\bf P} + {\bf q}/2$, and ${\bf p}_2 = -{\bf P} + {\bf q}/2$ and ${\bf p}_4 = -{\bf P} - {\bf q}/2$, respectively. Noting that ${\bf p}_n^2 = {\bf P}^2 + {\bf q}^2/4$, one sees that the initial and final energies of the fermions are roughly the same, that is, $q^0 \approx 0$ and $q^2 \approx -{\bf q}^2$, thus characterizing an elastic interaction. In the $\{ {\bf P}, {\bf q} \}$ basis, our approximations imply that terms of order higher than $\mathcal{O}(|{\bf P}|^2/m_f^2)$ and $\mathcal{O}(|{\bf q}|^2/m_f^2)$ will be neglected.

A direct consequence of our approximation scheme is that the NR normalization factors for the spinors, cf. Eq.~\eqref{eq_spinor}, are now mass and momentum dependent. The NR normalization is given by $N_{\rm NR} = \sqrt{(m_f + E)/2E}$, but up to second order we have $E \approx m_f \left( 1 + {\bf p}_n^2/2m_f^2 \right)$, therefore $N_{\rm NR} \rightarrow N_{\rm NR,f}$. For a given bilinear we always have the normalization factor squared, $N_{\rm NR,f}^2$, which may be expanded as (using ${\bf P}\cdot{\bf q} = 0$)
\begin{equation} \label{eq_app_NR_norm}
N_{\rm NR,f}^2 \approx 1 - \frac{1}{4m_f^2}\left(  {\bf P}^2 + \frac{{\bf q}^2}{4}  \right) \rightarrow  N_{\rm NR,a}^2 N_{\rm NR,b}^2 \approx 1 - \frac{1}{4} \left( \frac{1}{m_a^2} + \frac{1}{m_b^2} \right) \left(  {\bf P}^2 + \frac{{\bf q}^2}{4}  \right)  \, .
\end{equation}
This momentum-dependent pre-factor must be consistently accounted for in the calculation of the fermionic bilinears up to second order~\cite{Grupo2}.

\subsection{Fermionic bilinears} \label{app_B_0}
\indent

Here we explicitly calculate the fermionic bilinears that appear in the PV and T currents in the NR approximation. We use the standard Dirac representation of the gamma matrices
\begin{equation} \label{eq_gamma_matrices}
\gamma^0 = \left( \begin{array}{cc}
\mathbf{1} &  \,\,\mathbf{0} \\
\mathbf{0} & - \mathbf{1}   \end{array} \right) , \quad  \gamma^k = \left( \begin{array}{cc}
\,\,\mathbf{0} &  \sigma^k \\
- \sigma^k & \mathbf{0}   \end{array} \right) \quad  \textmd{and} \quad \gamma^5 = \left( \begin{array}{cc}
\,\,\mathbf{0} &  \,\,\mathbf{1}  \\
\,\,\mathbf{1}  & \mathbf{0}   \end{array} \right)  \, .
\end{equation}

The PV fermionic bilinear is $O^{\mu\nu}_{\rm PV} = \epsilon^{\mu\nu}_{\quad\beta\alpha} \bar{u}(p_f) \left(q^\alpha \gamma^\beta - q^\beta \gamma^\alpha \right)\gamma^5  u(p_i)$. In order to evaluate its components, it is worth quoting the identity
\begin{equation}
\bar{u}(p_f) \gamma^\mu \gamma^5  u(p_i) = \frac{1}{2m_f} \bar{u}(p_f) \left( q^\mu + 2iP_\nu \sigma^{\mu\nu} \right) \gamma^5 u(p_i) \, ,
\end{equation}
which can be obtained by using the Dirac equation and the fact that $\gamma^\mu \gamma^\nu = \eta^{\mu\nu} - i\sigma^{\mu\nu}$. Using these results we find that
\begin{equation} \label{eq_O_PV} 
O^{0k}_{\rm PV} = -4\epsilon_{klm} \left( {\bf p}_f - {\bf p}_i \right)_l \langle {\bm \sigma}_m \rangle \quad {\rm and } \quad O^{kl}_{\rm PV} = -\frac{2}{m_f} \epsilon_{klm} \left( {\bf p}_f - {\bf p}_i \right)_m \left[ \left( {\bf p}_i + {\bf p}_f \right) \cdot \langle {\bm \sigma} \rangle \right]  \, ,
\end{equation}
where we retained only terms of up to second order. In obtaining the results above we used $\sigma_i \sigma_j = \delta_{ij} + i\epsilon_{ijk}\sigma_k$ and $\sigma_i \sigma_j \sigma_l = \delta_{ij}\sigma_l - \delta_{il}\sigma_j + \delta_{jl}\sigma_i + i\epsilon_{ijl}$. For the T interaction the fermionic bilinear is $O^{\mu\nu}_{\rm T} = \bar{u}(p_f) \frac{i}{2}\left( \gamma^\mu \gamma^\nu - \gamma^\nu \gamma^\mu \right) u(p_i)$. Its components are given by
\begin{eqnarray}
O^{0k}_{\rm T} & = & -\frac{1}{2m_f} \epsilon_{klm} \left( {\bf p}_i + {\bf p}_f  \right)_l \langle {\bm \sigma}_m \rangle -\frac{i}{2m_f}\left( {\bf p}_f - {\bf p}_i  \right)_k  \, \label{eq_O_T_ok}  \\
O^{kl}_{\rm T} & = &  \epsilon_{klm} \Bigg\{  \langle {\bm \sigma}_m \rangle  - \frac{1}{4m_f^2} \left[ \left( {\bf p}_i \right)_m \left( {\bm \sigma} \cdot {\bf p}_f \right) + \left( {\bf p}_f \right)_m \left( {\bm \sigma} \cdot {\bf p}_i \right) - \left( {\bf p}_i \cdot {\bf p}_f \right) {\bm \sigma}_m \right]  \nonumber \\
& - &  \frac{i}{4m_f^2} \left[ \left( {\bf p}_i \right)_k \left( {\bf p}_f \right)_l - \left( {\bf p}_i \right)_l \left( {\bf p}_f \right)_k \right] \Bigg\} \, . \label{eq_O_T_kl} 
\end{eqnarray}
All fermionic bilinears, momentarily omitting the NR normalization, Eq.~\eqref{eq_app_NR_norm}, satisfy our limitation of keeping terms up to at most $\mathcal{O}(|{\bf P}|^2/m_f^2)$ and $\mathcal{O}(|{\bf q}|^2/m_f^2)$.

Note that the bilinears are antisymmetric, as expected, and we have omitted the factor $\xi_a^\dagger \xi_b = \delta_{ab}$ from the spinor normalization; this accounts for the impossibility of a spin flip. Moreover, the results above are 2-rank tensors, so that $O^{0k} = -O_{0k}$ and $O^{kl} = O_{kl}$.

\subsection{Non-relativistic amplitudes} \label{app_B_amps}
\indent

The NR PV amplitude for the KRLP-mediated interaction between two fermions labelled $a$ and $b$ is given by  (cf. Eqs.~\eqref{eq_vertex_PV} and~\eqref{eq_amp_H_1})
\begin{equation} \label{eq_app_amp_PV}
i\mathcal{M}_{\rm NR}^{\rm PV} = \left( \frac{ g_{\rm PV}^a g_{\rm PV}^b }{16\Lambda^2} \right) O_a^{\mu\nu} \Delta_{\mu\nu ,\alpha\beta}  O_b^{\alpha\beta} \, .
\end{equation}
The operators $O_{a,b}^{\mu\nu}$ are understood to represent the PV bilinears.  Due to the antisymmetry of the vertices, the terms in the propagator with four factors of the 4-momentum transfer automatically vanish and, after long, but straightforward manipulations, we are left with
\begin{equation}
\mathcal{M}_{\rm NR}^{\rm PV} = -\left( \frac{ g_{\rm PV}^a g_{\rm PV}^b }{16\Lambda^2} \right) \frac{1}{ {\bf q}^2 + m^2 } \left[ O_{a, \mu\nu} O_{b}^{\mu\nu} - \frac{2}{m^2} O_{a, \alpha\nu} O_{b}^{\alpha\beta} q^\nu q_\beta  \right] \, ,
\end{equation}
which may be further expanded using
\begin{eqnarray}
O_{a, \mu\nu} O_{b}^{\mu\nu} & = & - 2O_{a}^{0i} O_{b}^{0i} + O_{a}^{ij} O_{b}^{ij} \, , \\
O_{a, \alpha\nu} O_{b}^{\alpha\beta} q^\nu q_\beta & = & -{\bf q}_i {\bf q}_j \left( O_{a}^{ki} O_{b}^{kj} - O_{a}^{0i} O_{b}^{0j} \right) \, .
\end{eqnarray}

We evaluate the expressions above using Eq.~\eqref{eq_O_PV}. Keeping in mind that ${\bf q} \rightarrow -{\bf q}$ and ${\bf P} \rightarrow -{\bf P}$ in passing from the vertex for fermion $a$ to fermion $b$, we obtain
\begin{eqnarray}
\frac{1}{8} O_{a, \mu\nu} O_{b}^{\mu\nu} = {\bf q}^2 \left[ \langle {\bm \sigma} \rangle_a \cdot \langle {\bm \sigma} \rangle_b  + \frac{1}{m_a m_b} \left(  {\bf P} \cdot \langle {\bm \sigma} \rangle_a \right) \left(  {\bf P} \cdot \langle {\bm \sigma} \rangle_b \right)  \right] - \left(  {\bf q} \cdot \langle {\bm \sigma} \rangle_a \right) \left(  {\bf q} \cdot \langle {\bm \sigma} \rangle_b \right), 
\end{eqnarray} 
whereas $O_{a, \alpha\nu} O_{b}^{\alpha\beta} q^\nu q_\beta = 0$ due to symmetry. The NR PV amplitude at tree level is then
\begin{eqnarray} \label{eq_app_amp_PV_final}
\mathcal{M}_{\rm NR}^{\rm PV} & = & \left( \frac{ g_{\rm PV}^a g_{\rm PV}^b }{2 \Lambda^2} \right) \frac{1}{{\bf q}^2 + m^2} \Bigg\{ \left(  {\bf q} \cdot \langle {\bm \sigma} \rangle_a \right) \left(  {\bf q} \cdot \langle {\bm \sigma} \rangle_b \right) \nonumber \\
& -  & {\bf q}^2 \left[ \langle {\bm \sigma} \rangle_a \cdot \langle {\bm \sigma} \rangle_b  + \frac{1}{m_a m_b} \left(  {\bf P} \cdot \langle {\bm \sigma} \rangle_a \right) \left(  {\bf P} \cdot \langle {\bm \sigma} \rangle_b \right)  \right]  \Bigg\}  \, .
\end{eqnarray}
The NR T amplitude follows a similar structure as for the PV case, cf. Eq.~\eqref{eq_app_amp_PV}, namely
\begin{equation} \label{eq_app_amp_T}
\mathcal{M}_{\rm NR}^{\rm T} = -\frac{ g_{\rm T}^a g_{\rm T}^b }{ {\bf q}^2 + m^2 } \left[ O_{a, \mu\nu} O_{b}^{\mu\nu} - \frac{2}{m^2} O_{a, \alpha\nu} O_{b}^{\alpha\beta} q^\nu q_\beta  \right] \, 
\end{equation}
with the operators now referring to the T bilinears. Considering Eqs.~\eqref{eq_O_T_ok} and~\eqref{eq_O_T_kl} and keeping contributions only up to and including $\mathcal{O}(|{\bf P}|^2/m_f^2)$ and $\mathcal{O}(|{\bf q}|^2/m_f^2)$, we find
\begin{eqnarray} \label{eq_app_amp_T_final} 
\mathcal{M}_{\rm NR}^{\rm T} & = & \left( \frac{2g_{\rm T}^a g_{\rm T}^b}{m^2} \right) \left[ \langle {\bm \sigma} \rangle_a \cdot \langle {\bm \sigma} \rangle_b - \frac{1}{{\bf q}^2 + m^2} \left(  {\bf q} \cdot \langle {\bm \sigma} \rangle_a \right) \left(  {\bf q} \cdot \langle {\bm \sigma} \rangle_b \right)  \right] \nonumber \\
& - & \left( \frac{2g_{\rm T}^a g_{\rm T}^b}{m_a m_b} \right) \frac{1}{{\bf q}^2 + m^2}  \Bigg\{ \langle {\bm \sigma} \rangle_a \cdot \langle {\bm \sigma} \rangle_b \left( C_P {\bf P}^2 + C_q {\bf q}^2 \right) + \frac{1}{4}{\bf q}^2 \nonumber \\
& + & \left(  {\bf P} \cdot \langle {\bm \sigma} \rangle_a \right) \left(  {\bf P} \cdot \langle {\bm \sigma} \rangle_b \right) - \frac{i}{2}{\bf q} \cdot \left[ {\bf P} \times \left( \langle {\bm \sigma} \rangle_a + \langle {\bm \sigma} \rangle_b \right)  \right]  \Bigg\} \, ,  
\end{eqnarray} 
where we defined the coefficients
\begin{equation}
C_P = \frac{1}{4} \left( \frac{m_a}{m_b} + \frac{m_b}{m_a} - 4\right)  \quad {\rm \text{and}}  \quad C_q = \frac{1}{16} \left( \frac{m_a}{m_b} + \frac{m_b}{m_a}  \right)  \, .
\end{equation}

Similar to the PV amplitude, Eq.~\eqref{eq_app_amp_PV_final}, we have essentially a spin-spin interaction with small second-order corrections due to the velocity of the fermions. It is worth noting that Eq.~\eqref{eq_app_amp_T_final} presents an apparent divergence in the limit of a massless mediator, indicating that the interparticle potentials, as well as other quantities derived from this amplitude, will have an enhancement for light KRLPs. The velocity-dependent part of the NR T amplitude remains finite independently of $m$, but it is generally suppressed in a NR context.

\subsection{Relevant integrals} \label{app_ints}
\indent

Following Ref.~\cite{Maggiore}, the potential may be obtained by taking the Fourier transform of the NR amplitudes, as defined in Eq.~\eqref{eq_def_pot}. To do that the necessary integrals are
\begin{eqnarray}
\int \frac{d^3 \textbf{q}}{(2\pi)^3}  \, \frac{ 1 }{{\bf q}^2 + m^2} \,e^{i \textbf{q}\cdot\textbf{r}} & = & \frac{1}{r} y(r) \label{I_0} \, , \\
\int \frac{d^3 \textbf{q}}{(2\pi)^3}  \, \frac{ {\bf q}^{2} }{{\bf q}^2 + m^2} \,e^{i \textbf{q}\cdot\textbf{r}} & = & \delta^3({\bf r}) - \frac{m^2}{r} y(r) \label{I_1} \, , \label{eq_B17}\\
\int \frac{d^3 \textbf{q}}{(2\pi)^3}  \,   \frac{ {\bf q}_j }{{\bf q}^2 + m^2} \,e^{i \textbf{q}\cdot\textbf{r}} & = & \frac{i(1 + mr)}{r^2} \frac{r_j}{r} y(r) \label{I_1} \, , \\
\int \frac{d^3 \textbf{q}}{(2\pi)^3}  \,   \frac{ {\bf q}_i{\bf q}_j }{{\bf q}^2 + m^2} \,e^{i \textbf{q}\cdot\textbf{r}} & = &\frac{\delta_{ij}}{3} \delta^3(\textbf{r}) + \frac{1}{ r^3} \bigg[ (1 + m r) \delta_{ij} - (3 + 3 m r + m^2 r^2) \frac{r_i r_j}{r^2} \bigg] y(r) \label{I_ij} \, \label{eq_B19}
\end{eqnarray}
with $i,j = 1,2,3$ and $y(r) = e^{-m r}/4\pi$. The Dirac delta in Eq.~\eqref{I_ij} stems from the fact that Eq.~\eqref{eq_B17}$~= {\rm Tr}\left\{ {\rm Eq.~\eqref{eq_B19}} \right\}$~\cite{Dobrescu2, Ints}.

\section{Helicity eigenspinors: a brief overview} \label{app_helicity_spinors}
\indent

The evaluation of the unitarity bounds~\eqref{eq_unit_bounds} depends on the calculation of the amplitudes with all possible helicity states consistent with the total angular momentum $J$ of the partial wave under consideration. For this we must first find the basic spinors pointing at a generic direction $\hat{{\bf n}} = ( \sin\theta\cos\phi, \sin\theta\sin\phi, \cos\theta )$, where $\theta$ is the polar angle relative to the (positive) $z$-axis and $\phi$ is the azimuthal angle for a counter-clockwise rotation on the $x-y$ plane. We choose the coordinate system so that the initial electron-positron beam runs on the $z$-axis. It is therefore convenient to quantize the spin on the $z$-axis and take the eigenvectors of $\sigma_z$ as the basic spinors
\begin{equation} \label{eq_basis}
\xi_{+} = \left( \begin{array}{c} 1 \\ 0  \end{array} \right) \quad {\rm and} \quad \xi_{-} = \left( \begin{array}{c} 0 \\ 1  \end{array} \right) \, .
\end{equation}

\newpage
We must now rotate $\xi_{\pm}$ toward $\hat{{\bf n}}$ by means of the matrix $U(\omega) = \exp\left( -i \omega_{\mu\nu} S^{\mu\nu}/2 \right)$, where $S^{\mu\nu} = (i/4) \left[ \gamma^\mu, \gamma^\nu \right]$ are the generators of Lorentz boosts and spatial rotations. Starting at $\theta = \phi = 0$ we first rotate around the $y$-axis by an angle $\omega_{31} = -\omega_{13} = \theta$ and then around the $z$-axis by an angle $\omega_{12} = -\omega_{21} = \phi$. The unitary operator that accomplishes this is $U(\theta, \phi) = e^{-i(\phi/2)\sigma_z} e^{-i(\theta/2)\sigma_y}$, which can be explicitly written as
\begin{equation} \label{eq_U}
U(\theta, \phi) = e^{-i\phi/2} \left( \begin{array}{cc}
\cos\frac{\theta}{2}              & -\sin\frac{\theta}{2} \\
e^{i\phi}\sin\frac{\theta}{2}  & e^{i\phi}\cos\frac{\theta}{2}  \end{array} \right)  \, .
\end{equation}
The overall phase factor is inconsequential and will be henceforth omitted. The basic spinors along $\hat{{\bf n}}$ are then $U(\theta, \phi) \xi_\pm = \xi'_\pm$, namely
\begin{equation} \label{eq_basis_rot}
\xi'_{+} = \left( \begin{array}{c} \cos\frac{\theta}{2} \\ e^{i\phi}\sin\frac{\theta}{2}  \end{array} \right) \quad {\rm and} \quad \xi'_{-} = \left( \begin{array}{c} -\sin\frac{\theta}{2} \\ e^{i\phi}\cos\frac{\theta}{2}  \end{array} \right) \, .
\end{equation}
As expected, these basic spinors also form an orthonormal basis.

Again starting from a particle moving in the positive $z$-direction described by the spinor~\eqref{eq_spinor} with ${\bf p} = (0,0,|{\bf p}|)$ we find the associated rotated spinor by applying $U(\theta, \phi)$ to the upper and lower components. For the upper component we get simply $\xi'_\pm$, whereas the lower component is proportional to $U(\theta, \phi) \sigma_z \xi_\pm$. However, we know that the basic spinors $\xi_\pm$ are eigenvectors of $\sigma_z$ with $\sigma_z \xi_\pm = \pm \xi_\pm$, so that the lower component turns out to be $\frac{\pm | {\bf p} |}{E + m_f} \xi'_\pm$. This makes sense, since the rotated spinors are eigenvectors of ${\bm \sigma} \cdot \left( |{\bf p}| \hat{ {\bf n} } \right)$. A similar procedure must be applied to the spinor for anti-particles. Therefore, the spinors for particles and anti-particles moving with 3-momentum of magnitude $|{\bf p}|$ in a generic direction $\hat{ {\bf n} }$ are
\begin{equation}\label{eq_spinors_u_v}
u^\pm (p) = N_{\rm R} \left( \begin{array}{c}
\xi'_\pm  \\   \frac{ \pm |{\bf p}| }{E+m_f} \xi'_\pm  \end{array} \right) \quad {\rm and} \quad v^\pm (p) = \pm N_{\rm R} \left( \begin{array}{c}
\frac{ \mp |{\bf p}|  }{E+m_f} \xi'_\mp  \\  \xi'_\mp  \end{array} \right) \, 
\end{equation}
with $N_{\rm R} = \sqrt{E + m_f}$. These spiunors satisfy $u_\pm^\dagger u_\pm = v_\pm^\dagger v_\pm = 2E$ and $\bar{u}_\pm u_\pm = -\bar{v}_\pm v_\pm = 2m$; all other combinations are zero.

The helicity operator, given by $\hat{h} = {\bm \Sigma}\cdot \hat{{\bf p}}$ with ${\bm \Sigma} = {\rm diag} ({\bm \sigma}, {\bm \sigma})$ and $\hat{{\bf p}} = {\bf p}/|{\bf p}| = \hat{{\bf n}}$, projects the spin along the direction of motion. Noting that $\left( {\bm \sigma}\cdot \hat{{\bf n}} \right) \xi'_\pm = \pm \xi'_\pm$ and keeping in mind that the helicity operator applied to anti-particles is defined as $\hat{h}_v \equiv -\hat{h}$, we find
\begin{equation} \label{eq_def_helicity}
\hat{h} u^\pm (p) = \pm u^\pm (p)  \quad {\rm and} \quad \hat{h}_v v^\pm (p) = \pm v^\pm (p) \, .
\end{equation} 
The subscripts $\pm$ label the helicity eigenvalues of the spinors, $\pm 1$. Finally, in the ultra-relativist regime, where $E = \sqrt{s}/2 \gg m_f$, the spinors~\eqref{eq_spinors_u_v} simplify to 
\begin{equation}\label{eq_spinors_u_v_UR}
u^\pm (p) = \sqrt{E} \left( \begin{array}{c}
\xi'_\pm  \\   \pm \xi'_\pm  \end{array} \right) \quad {\rm and} \quad v^\pm (p) = \pm \sqrt{E} \left( \begin{array}{c}
\mp \xi'_\mp  \\  \xi'_\mp  \end{array} \right) \, .
\end{equation}

\section{The KRLP-mediated differential cross sections} \label{app_dcs}
\indent

In Sec.~\eqref{sec_kinematics} we stated the $s$- and $t$-channel amplitudes for $2 \rightarrow 2$ lepton scattering in the CM frame. Assuming $s = E_{\rm CM}^2 \gg m_f^2$, the unpolarized differential cross section is~\cite{Maggiore}
\begin{equation}
\frac{d \sigma}{d\Omega} = \frac{1}{64\pi^2 s} \langle |\mathcal{M}|^2  \rangle \, ,
\end{equation}
where $d\Omega = d\cos\theta d\phi$ and $\phi = \left[ 0, 2\pi \right]$. Here $\langle |\mathcal{M}|^2  \rangle$ indicates averaging the amplitude squared over initial spins and summing over final spins; typically the result is independent of the azimuthal angle $\phi$. If the amplitude is $\mathcal{M} = \mathcal{M}_1 + \mathcal{M}_2$ we will have ``pure" contributions $\sim |\mathcal{M}_{1,2}|^2$, but also so-called interference terms  $\sim |\mathcal{M}_{1}^\star \mathcal{M}_{2}|^2$.

For the KRLP scenario with PV vertices we find the following interference contributions with photon-mediated diagrams
\begin{eqnarray} \label{eq_dcs_PV_qed} 
\frac{d \sigma^{\rm PV}_{\gamma}}{d(\cos\theta)} & = & \frac{ \alpha g_{\rm PV}^{e} g_{\rm PV}^{e} }{ 16\Lambda^2} \frac{ y^2 }{ \left( 1 - x \right) \left( 1 - y^2\right) \left[2 +(1- x)  y^2  \right]}  \\
& \times & \left[  (11 - 10 x + 12 x^2 + 2 x^3 + x^4) - y^2 (5 - 8 x - 6 x^2 + 8 x^3 + x^4)  \right] \nonumber
\end{eqnarray}
and with Z-mediated diagrams
\begin{eqnarray}\label{eq_dcs_PV_EW} 
\frac{d \sigma^{\rm PV}_{\rm Z}}{d(\cos\theta)} & = & \frac{ g_{\rm PV}^{e} g_{\rm PV}^{e} }{ 32\pi \Lambda^2}
  \frac{ g_Z^2 y^2 z^2}{ \left( 1 - y^2\right) \left[2 +(1- x)  y^2  \right]}  \\
& \times &  \frac{1}{ \left[ \left(1- z^2 \right)^2+w^2 \right] \left[ \left(z^2 (1- x) +2\right)^2 + 4 w^2\right]} \nonumber \\
& \times & \Bigg\{  (g_V^2-g_A^2)\Bigg[(1 + w^2) (1 - x^2) - z^2 \Big( (3 - 3 x + x^2 - x^3) - 2w^2(1 - 2 x + x^2) \Big) \nonumber \\
&  + & \dfrac{1}{4}z^4 (3 - 2 x^2 - x^4) +\dfrac{1}{4}z^6 (5 - 12 x + 10 x^2 - 4 x^3 + x^4) \nonumber \\ 
& - & \dfrac{1}{2} y^2 \bigg( (1+w^2)(5 - 7 x + 3 x^2 - x^3) \nonumber \\ 
& - & z^2 \Big( (6 - 3 x - 5 x^2 + 3 x^3 - x^4) - w^2 (1 - x - x^2 + x^3) \Big) \nonumber \\ 
& - & \dfrac{1}{4}z^4(3 - 27 x + 30 x^2 - 6 x^3 - x^4 + x^5) + \dfrac{1}{4}z^6 (7 - 11 x - 2 x^2 + 10 x^3 - 5 x^4 + x^5)\bigg)  \Bigg] \nonumber \\
& + & (g_V^2+g_A^2)\Bigg[ (1 + w^2)(x^3 + 3 x^2 + 3 x + 1) \nonumber \\
& - & \dfrac{1}{4}z^2 \Big( 3 (1 + 4 x + 6 x^2 + 4 x^3 + x^4) + w^2 (1 + 4 x + 6 x^2 + 4 x^3 + x^4) \Big)\nonumber \\ 
& -& \dfrac{1}{8}z^4 (3 + 3 x - 10 x^2 - 18 x^3 - 9 x^4 - x^5) + \dfrac{1}{8} z^6 (1 + 3 x + 2 x^2 - 2 x^3 - 3 x^4 - x^5) \nonumber \\
& + & 2 y^2 \bigg((1+w^2)(1 + x - x^2 - x^3) \nonumber \\
& - & \dfrac{1}{4}z^2 \Big( 3 (1 + 2 x - 2 x^3 - x^4) + w^2 (1 + 2 x - 2 x^3 - x^4)\Big)  \bigg) \nonumber \\
& - & \dfrac{1}{8} z^4(3 - 3 x - 10 x^2 + 2 x^3 + 7 x^4 + x^5) +\dfrac{1}{8} z^6 (1 + x - 2 x^2 - 2 x^3 + x^4 + x^5)\Bigg] 
 \Bigg\} \, ,   \nonumber 
\end{eqnarray}
where $x = \cos\theta$, $y =\sqrt{s}/m$, $z = \sqrt{s}/m_Z$ and $w = \Gamma_Z/m_Z$. Note the resonance at $y = 1$ ($\sqrt{s} = m$), as well as the forward divergence as $x \rightarrow 1$ inherited from QED. The pure PV diffferential cross section is
\begin{eqnarray}\label{eq_dcs_PV} 
\frac{d \sigma^{\rm PV}}{d(\cos\theta)} & = & \left( \frac{g_{\rm PV}^{e} g_{\rm PV}^{e}}{\Lambda^2} \right)^2 \frac{m^2}{256\pi} \frac{y^6}{  \left( 1 - y^2 \right)^2 \left[ 2 + \left( 1 - x \right)y^2 \right]^2 } \nonumber \\
& \times & \Bigg[ (9 - 12 x + 14 x^2 + 4 x^3 + x^4) + 4y^2 (3x + x^2 - 3 x^3 - x^4) \nonumber \\
& + & y^4 (1 - x)^2 (9 + 6 x + 5 x^2)  \Bigg]   \, .
\end{eqnarray}

The leading-order contribution from KRLPs with T vertices stems from interference with pure QED:
\begin{eqnarray}\label{eq_dcs_T_qed} 
\frac{d \sigma_{\gamma}^{\rm T}}{d(\cos\theta)} & = & -\frac{\alpha g_{\rm T}^{e} g_{\rm T}^{e} }{8m^2} \frac{y^2}{ \left( 1 - x \right) \left( 1 - y^2 \right) \left[ 2 + \left( 1 - x \right)y^2 \right] } \nonumber \\
& \times & \left[ \left( 3 - 24x + 6x^2 - x^4 \right)  - y^2 \left( 3 - 2x^2 - x^4  \right)  \right] \, , 
\end{eqnarray}
where again the resonance at $y = 1$ ($\sqrt{s} = m$) and the forward divergence at $x \rightarrow 1$ are present. The interference with Z-mediated diagrams is
\begin{eqnarray} \label{eq_dcs_T_EW} 
\frac{d \sigma_{\rm Z}^{\rm T}}{d(\cos\theta)} & = & \frac{ g_{\rm T}^{e} g_{\rm T}^{e}  }{32\pi m^2} \frac{  g_Z^2 (g_V^2-g_A^2) y^2 z^2 }{ \left( 1 - y^2\right) \left[2 +(1- x)  y^2  \right]}  \nonumber \\
& \times & \frac{1 }{ \left[ \left(1- z^2 \right)^2+w^2 \right] \left[ \left(z^2 (1- x) +2\right)^2 + 4w^2\right]}  \Bigg\{ (1 + w^2) (3 + 3 x + x^2 + x^3)  \nonumber \\
& - & z^2 \bigg[ x(15  - x + x^2 + x^3) + w^2 (3 - 9 x + 5 x^2 + x^3) \nonumber \\
& - & \dfrac{1}{4} z^4(9 - 21 x - 22 x^2 + 6 x^3 - 3 x^4 - x^5) - \dfrac{1}{4} z^6 (3 - 27x + 30x^2 - 6x^3 - x^4 + x^5 )\bigg] \nonumber \\
& - & y^2 \bigg[ (1 + w^2) (3 - 9 x + 5 x^2 + x^3) \nonumber \\
& + & z^2 \Big(x (3  + x - 3 x^2 - x^3 ) - 3 w^2 (1 - x - x^2 + x^3) \Big) \nonumber \\ 
& - & \dfrac{1}{4} z^4 (9 - 21 x + 26 x^2 - 10 x^3 - 3 x^4 - x^5)  \nonumber \\
& - & \dfrac{1}{4}z^6 (3 - 3 x - 2 x^2 + 2 x^3 - x^4 + x^5) \bigg] \Bigg\} \, .
\end{eqnarray}
Finally, we also have the pure T contribution:
\begin{eqnarray} \label{eq_dcs_T} 
\frac{d \sigma^{\rm T}}{d(\cos\theta)} & = & \frac{ \left( g_{\rm T}^{e} g_{\rm T}^{e} \right)^2  }{4\pi m^2} \frac{ y^2 }{ \left( 1 - y^2 \right)^2 \left[ 2 + \left( 1 - x \right) y^2 \right]^2 } \nonumber \\
& \times & \bigg[ 9 \left( 1 + x \right)^2 - \frac{9}{2} y^2 (5+3x-x^2+x^3)  \nonumber \\
& + & \frac{1}{8} y^4 (153-60x+34x^2-52x^3+5x^4)   \nonumber \\
& + & \frac{1}{4} y^6 (27+15x+23x^2-15x^3+4x^4)     \nonumber \\
& + & \frac{1}{8} y^8 (1-x)^2(9+12x+7x^2)  \bigg] \, .
\end{eqnarray}


{\onehalfspacing

}


\begin{thebibliography}{99}


\bibitem{notoph1}
V.~I.~Ogievetsky and I.~V.~Polubarinov,
\textit{The notoph and its possible interactions},
Yad. Fiz. \textbf{4}, 216 (1966)  [Sov. J. Nucl. Phys. {\bf 4}, 156 (1967)].

\bibitem{notoph2}
E.~A.~Ivanov,
\textit{Gauge Fields, Nonlinear Realizations, Supersymmetry},
Phys. Part. Nucl. \textbf{47},  508 (2016). 

\bibitem{KR} M. Kalb, P. Ramond, {\it Classical Direct Interstring Action}, Phys. Rev. D {\bf 9}, 2273 (1974).

\bibitem{Cremmer} E. Cremmer, J. Scherk,
\textit{Spontaneous dynamical breaking of gauge symmetry in dual models}, Nuclear Physics \textbf{72}, 117-124 (1974).



\bibitem{Becker}
K.~Becker, M.~Becker and J.~H.~Schwarz,
\textit{String theory and M-theory: A modern introduction},
Cambridge University Press, (2006)


\bibitem{Dvali} I. Antoniadis, N. Arkani-Hamed, S. Dimopoulos,
G.R. Dvali, {\it New dimensions at a millimeter to a fermi
and superstrings at a TeV}, Phys. Lett. B {\bf 436}, 257 (1998).

\bibitem{Dick0} S. Tiwary, R. Dick, {\it Constraints on antisymmetric tensor fields from Bhabha scattering}, Eur. Phys. J. C {\bf 81}, 1115 (2021).

\bibitem{EW_KR} A. Dashko, R. Dick, {\it The shadow of dark matter as a shadow of string theory}, Eur. Phys. J. C {\bf 79}, 312 (2019).

\bibitem{Cline} J.M. Cline, G.D. Moore, A.R. Frey, {\it Composite magnetic dark matter and the 130 GeV line}, Phys. Rev. D {\bf 86}, 115013 (2012).

\bibitem{Dick} A. J. Magnus, J. G. Fenwick, R. Dick, {\it Antisymmetric tensor portals to dark matter}, arXiv:hep-ph/2409.05915.

\bibitem{NambuKR} Y.~Nambu, \textit{III. Magnetic and electric confinement of quarks} Physics Reports \textbf{23}(3), 250–253 (1976).

\bibitem{greensite}
J.~Greensite, {\it An Introduction to Confinement Problem}, Lecture Notes in Physics \textbf{821}, second edition (Springer, Berlin, 2020).

\bibitem{Grigorio:2011pi}
L.~S.~Grigorio, M.~S.~Guimaraes, R.~Rougemont and C.~Wotzasek,
{\it Confinement, brane symmetry and the Julia-Toulouse approach for condensation of defects},
JHEP \textbf{08}, 118 (2011).

\bibitem{Guimaraes:2012tx}
M.~S.~Guimaraes, R.~Rougemont, C.~Wotzasek and C.~A.~D.~Zarro,
{\it Massive photons and Dirac monopoles: electric condensate and magnetic confinement},
Phys. Lett. B \textbf{723}, 422-426 (2013).

\bibitem{Smailagic:2020kep}
A.~Smailagic and E.~Spallucci,
{\it Cornell potential in Kalb-Ramond scalar QED via Higgs mechanism},
Phys. Lett. B \textbf{803}, 135304 (2020).

\bibitem{Franz:2007} M. Franz, \textit{Vortex-boson duality in four space-time dimensions}, EPL, \textbf{77}(4), 47005 (2007).  

\bibitem{Rougemont:2015gia}
R.~Rougemont, J.~Noronha, C.~A.~D.~Zarro, C.~Wotzasek, M.~S.~Guimaraes and D.~R.~Granado,
{\it Vanishing DC holographic conductivity from a magnetic monopole condensate},
JHEP \textbf{07}, 070 (2015).

\bibitem{Barone:2005hn}
F.~A.~Barone, L.~M.~De Moraes and J.~A.~Helayel-Neto,
{\it Casimir effect for gauge scalars: The Kalb-Ramond case},
Phys. Rev. D \textbf{72}, 105012 (2005)
[erratum: Phys. Rev. D \textbf{73}, 089901 (2006)].

\bibitem{Belich:2010xj}
H.~Belich, L.~M.~Silva, J.~A.~Helayel-Neto and A.~E.~Santana,
{\it Casimir Effect at finite temperature for the Kalb-Ramond field},
Phys. Rev. D \textbf{84}, 045007 (2011).

\bibitem{Smailagic:2021poa}
A.~Smailagic and E.~Spallucci,
{\it Kalb\textendash{}Ramond scalar QED multiple vacua},
J. Phys. G \textbf{48}, no.12, 125002 (2021).



\bibitem{Barone:2010zi}
F.~A.~Barone, F.~E.~Barone and J.~A.~Helayel-Neto,
{\it Charged branes interactions via Kalb-Ramond field},
Phys. Rev. D \textbf{84}, 065026 (2011).

\bibitem{Petrov1} J. F. Assun\c{c}\~ao, T. Mariz, J. R. Nascimento, A. Yu. Petrov, {\it Dynamical Lorentz symmetry breaking in a tensor bumblebee model}, Phys. Rev. D {\bf 100}, 085009 (2019).

\bibitem{Petrov2} L. H. C. Borges, A. G. Dias, A. F. Ferrari, J. R. Nascimento, A. Yu. Petrov, {\it Generation of Axion-Like Couplings via Quantum Corrections in a Lorentz Violating Background}, Phys. Rev. D {\bf 89}, 045005 (2014).

\bibitem{Petrov3} A. J. G. Carvalho, A. G. Dias, A. F. Ferrari, T. Mariz, J. R. Nascimento, A. Yu. Petrov, {\it Axion-Photon Interaction from Nonminimal Dimension-5 Lorentz-Violating Operators}, Phys. Rev. D {\bf 107}, 085021 (2023).

\bibitem{Ferreira:2008yf}
C.~N.~Ferreira, J.~A.~Helayel-Neto and N.~A.~Tomimura,
{\it Plane Gravitational Radiation from Neutrinos Source with Kalb-Ramond Coupling},
Int. J. Mod. Phys. A \textbf{24}, 1537-1540 (2009).

\bibitem{Chakraborty:2017uku}
S.~Chakraborty and S.~Sengupta,
{\it Packing extra mass in compact stellar structures: An interplay between Kalb-Ramond field and extra dimensions},
JCAP \textbf{05}, 032 (2018)

\bibitem{Junior:2024ety}
E.~L.~B.~Junior, J.~T.~S.~S.~Junior, F.~S.~N.~Lobo, M.~E.~Rodrigues, D.~Rubiera-Garcia, L.~F.~D.~da Silva and H.~A.~Vieira,
{\it Spontaneous Lorentz symmetry-breaking constraints in Kalb\textendash{}Ramond gravity},
Eur. Phys. J. C \textbf{84}, no.12, 1257 (2024).

\bibitem{Cosm_1} D. Maity, P. Majumdar, S. SenGupta, {\it Parity violating Kalb-Ramond-Maxwell interactions and CMB anisotropy in a brane world}, JCAP {\bf 06}, 005 (2004).

\bibitem{Cosm_2} J. P. Beltr\'an Almeida, A. Guarnizo and C. A. Valenzuela-Toledo, {\it Arbitrarily coupled
p-forms in cosmological backgrounds}, Class. Quant. Grav. {\bf 37}, 035001 (2020).

\bibitem{Cosm_3} J. P. B. Almeida, A. Guarnizo, R. Kase, S. Tsujikawa, C. A. Valenzuela-Toledo, {\it Arbitrarily coupled p-forms in cosmological backgrounds}, JCAP {\bf 03}, 025 (2019).

\bibitem{Capanelli:2023uwv}
C.~Capanelli, L.~Jenks, E.~W.~Kolb and E.~McDonough,
{\it Cosmological implications of Kalb-Ramond-like particles},
JHEP \textbf{06}, 075 (2024).

\bibitem{dual} S. E. Hjelmeland, U. Lindstrom, {\it Duality for the nonspecialist}, arXiv:hep-th/9705122.

\bibitem{Hell} A. Hell, {\it On the duality of massive Kalb-Ramond and Proca fields}, JCAP {\bf 01}, 056 (2022).

\bibitem{Witten} P. Svrcek, E. Witten, {\it Axions In String Theory}, JHEP {\bf 06}, 051 (2006).

\bibitem{Russell} A. Arvanitaki, S. Dimopoulos, S. Dubovsky, N. Kaloper, and J. March-Russell, {\it String Axiverse}, Phys. Rev. D {\bf 81}, 123530 (2010).

\bibitem{PQ} R.D. Peccei, H.R. Quinn, {\it CP Conservation in the Presence of Instantons}, Phys. Rev. Lett. {\b 38}, 1440 (1977).

\bibitem{Weinberg} S. Weinberg, {\it A New Light Boson?}, Phys. Rev. Lett. {\bf 40}, 223 (1978).

\bibitem{Wilczek} F. Wilczek, {\it Problem of Strong P and T Invariance in the Presence of Instantons}, Phys. Rev. Lett. {\bf 40}, 279 (1978).



\bibitem{Okun} L.B. Okun, Limits of electrodynamics: paraphotons?, Sov. Phys. JETP {\bf 56}, 502 (1982).

\bibitem{Holdon} B. Holdom, Two U(1)'s and epsilon charge shifts, Phys. Lett. B {\bf 166}, 196 (1986).

\bibitem{Abel1} S.A. Abel, M.D. Goodsell, J. Jaeckel, V.V. Khoze, A. Ringwald, Kinetic mixing of the photon with hidden U(1)s in string phenomenology, JHEP {\bf 807}, 124 (2008).

\bibitem{Abel2} S. Abel, J. Santiago, Constraining the string scale: from Planck to Weak and back again, J. Phys. G {\bf 30}, (2004).

\bibitem{Fayet1} P. Fayet, Extra U(1)'s and new forces, Nucl. Phys. B {\bf 347}, 743 (1990).

\bibitem{Smailagic}
A. Smailagic and E. Spallucci, {\it The dual phases of massless/massive Kalb-Ramond fields}, J. Phys. A: Math. Gen. \textbf{34} L435 (2001).

\bibitem{DM1} K.M. Zurek, {\it Dark Matter Candidates of a Very Low Mass}, arXiv:hep/ph:2401.03025.

\bibitem{DM2} J.L. Feng {\it Dark Matter Candidates from Particle Physics and Methods of Detection}, Ann. Rev. Astron. Astrophys. {\bf 48}, 495 (2010).

\bibitem{DM3} F. Chadha-Day, J. Ellis, D.J.E. Marsh, {\it Axion Dark Matter: What is it and Why Now?}, arXiv:hep-ph/2105.01406.

\bibitem{DM4} A.E. Nelson, J. Scholtz, {\it Dark Light, Dark Matter and the Misalignment Mechanism}, Phys. Rev. D {\bf 84}, 103501 (2011).

\bibitem{DM5} E. Mass\'o, F. Rota, G. Zsembinszki, {\it Planck-Scale Effects on Global Symmetries: Cosmology of Pseudo-Goldstone Bosons}, Phys. Rev. D {\bf 70}, 115009 (2004).

\bibitem{Graham} P.W. Graham {\it et al.}, {\it Experimental Searches for the Axion and Axion-Like Particles}, Ann. Rev. Nucl. Part. Sci. {\bf 65}, 485 (2015).

\bibitem{Capozzi} F. Capozzi {\it et al.}, {\it New Constraints on ALP Electron and Photon Couplings from ArgoNeuT and the MiniBooNE Beam Dump}, arXiv:hep-ph/2307.03878.

\bibitem{FIPS} P. Agrawal {\it et al.}, {\it Feebly-interacting particles: FIPs 2020 Workshop Report (2021)}, arXiv:hep-ph/2102.12143.

\bibitem{JJ} J. Jaeckel, {\it A force beyond the Standard Model - Status of the quest for hidden photons}, arXiv:hep-ph/1303.1821.

\bibitem{Leike} A. Leike, {\it The phenomenology of extra neutral gauge bosons}, Phys. Rept. {\bf 317}, 143 (1999).

\bibitem{Langacker} P. Langacker, {\it The physics of heavy $Z'$ gauge bosons}, Rev. Mod. Phys. {\bf 81}, 1199 (2009).








\bibitem{feyncalc1}
V. Shtabovenko, R. Mertig and F. Orellana, {\it FeynCalc 10: Do multiloop integrals dream of computer codes?}, arXiv:2312.14089.

\bibitem{feyncalc2}
V. Shtabovenko, R. Mertig and F. Orellana, {\it FeynCalc 9.3: New features and improvements}, arXiv:2001.04407.

\bibitem{feyncalc3}
V. Shtabovenko, R. Mertig and F. Orellana, {\it New Developments in FeynCalc 9.0}, Comput. Phys. Commun. \textbf{207}, 432-444, (2016).

\bibitem{feyncalc4}
R. Mertig, M. Böhm, and A. Denner, {\it Feyn Calc - Computer-algebraic calculation of Feynman amplitudes}, Comput. Phys. Commun. \textbf{64}, 345-359  (1991).


















\bibitem{Liu} J. Liu, Y. Luo, M. Song, {\it Investigation of the concurrent effects of ALP-photon and ALP-electron couplings in Collider and Beam Dump Searches}, JHEP {\bf 09}, 104 (2023).

\bibitem{Nambu} Y. Nambu, {\it Axial vector current conservation in weak interactions}, Phys. Rev. Lett. {\bf 4}, 380 (1960).

\bibitem{Cao} X.-H. Cao, Z.-H. Guo, {\it Comprehensive study of axion photoproduction off the nucleon in chiral effective field theory}, Phys. Rev. D {\bf 110}, 095025 (2024).

\bibitem{Naydenov} M. Chizhov, M. Naydenov, {\it Isospin-invariant Nambu–Jona-Lasinio model with complete set of spin-1 excitations},
AIP Conf. Proc. {\bf 2075}, 090025 (2019).

\bibitem{tensor1} M. Naydenov, V. Kozhuharov, {\it Dark boson mediation of the $\pi^0 \rightarrow \gamma e^+ e^-$ decay}, Nucl. Phys. B {\bf 978}, 115723 (2022).

\bibitem{tensor2} M. Naydenov, V. Kozhuharov, {\it Dark sector tensor currents contribution to lepton's anomalous magnetic moment}, arXiv:hep-ph/2212.02242.

\bibitem{Dobrescu} B.A. Dobrescu, {\it Massless Gauge Bosons other than the Photon}, Phys. Rev. Lett. {\bf 94}, 151802 (2005).


\bibitem{Bailey} B. Altschul, Q.G. Bailey, V.A. Kostelecky, {\it Lorentz violation with an antisymmetric tensor}, Phys. Rev. D {\bf 81}, 065028 (2010).






\bibitem{Maggiore} M. Maggiore, \textit{A modern introduction to quantum field theory}, Oxford University Press (2005).

\bibitem{Grupo1} F.A. Gomes Ferreira, P.C. Malta, L.P.R. Ospedal, J.A. Helay\"el-Neto, {\it Topologically massive spin-1 particles and spin-dependent potentials}, Eur. Phys. J. C {\bf 75}, 238 (2015).

\bibitem{Grupo2} P.C. Malta, L.P.R. Ospedal, K. Veiga, J.A. Helay\"el-Neto, {\it Comparative aspects of spin-dependent interaction potentials for spin-1/2 and spin-1 matter fields}, Adv. High Energy Phys. 2016 (2016) 2531436, Adv. High Energy Phys. 2017 (2017) 9152437 (erratum).

\bibitem{Grupo3} G.P. de Brito, P.C. Malta, L.P.R. Ospedal, {\it Spin- and velocity-dependent nonrelativistic potentials in modified electrodynamics}, Phys. Rev. D {\bf 95}, 016006 (2017).

\bibitem{Fadeev} P. Fadeev, Y.V. Stadnik, F. Ficek, M.G. Kozlov, V.V. Flambaum, D. Budker, {\it Revisiting spin-dependent forces mediated by new bosons: Potentials in the coordinate-space representation for macroscopic- and atomic-scale experiments}, Phys. Rev. A {\bf 99},
022113 (2019).

\bibitem{Fadeev2} P. Fadeev, F. Ficek, M.G. Kozlov, D. Budker, V.V. Flambaum, {\it Pseudovector and pseudoscalar spin-dependent interactions in atoms}, Phys. Rev. A {\bf 105},
022812 (2022).

\bibitem{Cong} L. Cong, F. Ficek, P. Fadeev, D. Budker, {\it Improved constraints on exotic interactions between electron and proton in hydrogen}, arXiv:physics.atom-ph/2408.11009.

\bibitem{Cong2} L. Cong  {\it et al.}, {\it Spin-dependent exotic interactions}, arXiv:hep-ph/2408.15691.

\bibitem{Dobrescu2} B.A. Dobrescu, I. Mocioiu, {\it Spin-Dependent Macroscopic Forces from New Particle Exchange}, 	JHEP {\bf 0611}, 005 (2006).

\bibitem{Joao} J. P. S. Melo, M. J. Neves, J. M. A. Paixão, J. A. Helay\"{e}l-Neto. Loop quantum gravity effects on electromagnetic properties of charged leptons. Eur. Phys. J. C \textbf{84}, 938 (2024).

\bibitem{Cohen_book} Claude Cohen-Tannoudji, Bernard Diu, Franck Lalo\"e, {\it Quantum Mechanics (vol. I, II)}, 2nd edition, Wiley (1977).

\bibitem{Griffiths_hyp} D. Griffiths, {\it Hyperfine Splitting in the ground state of Hydrogen}, Am. J. Phys. {\bf 50}, 698 (1982).

\bibitem{Parthey} N. Kolachevsky, A. Matveev, J. Alnis, C.G. Parthey, S.G. Karshenboim, T.W. Haensch, {\it New Measurement of the 2S Hyperfine Interval in Atomic Hydrogen}, Phys. Rev. Lett. {\bf 102}, 213002 (2009).

\bibitem{Parthey2} C.G. Parthey {\it et al.}, {\it Improved Measurement of the Hydrogen 1S - 2S Transition Frequency}, Phys. Rev. Lett. {\bf 107}, 203001 (2011).

\bibitem{Bullis_hfs} R. G. Bullis, C. Rasor, W. L. Tavis, S. A. Johnson, M. R. Weiss, D. C. Yost, {\it Ramsey Spectroscopy of the $2S_{1/2}$ Hyperfine Interval in Atomic Hydrogen}, Phys. Rev. Lett. {\bf 130}, 203001 (2023).

\bibitem{D21} M.M. Sternheim, {\it State-Dependent Mass Corrections to Hyperfine Structure in Hydrogenic Atoms}, Phys. Rev. {\bf 130}, 211 (1963).

\bibitem{Ficek} F. Ficek, D. Budker, {\it Constraining Exotic Interactions}, Annalen der Physik {\bf 531}, 1800273 (2019).

\bibitem{D21_th} V.A. Yerokhin, U.D. Jentschura, {\it Electron Self-Energy in the Presence of a Magnetic Field: Hyperfine Splitting and g Factor}, Phys. Rev. Lett. {\bf 100}, 163001 (2008).


\bibitem{Joerg_HP} J. Jaeckel, S. Roy, {\it Spectroscopy as a test of Coulomb's law - A probe of the hidden sector}, Phys. Rev. D {\bf 82}, 125020 (2010).


\bibitem{Kroff_Malta} D. Kroff, P.C. Malta, {\it Constraining hidden photons via atomic force microscope measurements and the Plimpton-Lawton experiment}, Phys. Rev. D {\bf 102}, 095015 (2020).














\bibitem{LEP} G. Montagna, O. Nicrosini, F. Piccinini, {\it Precision physics at LEP}, Riv. Nuovo Cim. 21N9, 1-162 (1998).

\bibitem{SLD} SLD collaboration, {\it Polarized Bhabha scattering and a precision measurement of the electron neutral current couplings}, Phys. Rev. Lett. {\bf 74}, 2880 (1995).  

\bibitem{Likic} I. Bozovic-Jelisavcic, S. Lukic, M. Pandurovic, I. Smiljanic, {\it Precision luminosity measurement at ILC}; arXiv:physics.acc-ph/1403.7348v1.

\bibitem{Bicer} M. Bicer {\it et al.}, TLEP Design Study Working Group, JHEP {\bf 01}, 164 (2014).

\bibitem{Baer} H. Baer {\it et al.}, {\it The International Linear Collider Technical Design Report -- Volume 2: Physics}, arXiv:hep-ph/1306.6352.

\bibitem{Derrick} M. Derrick \textit{et al.}, \textit{Experimental study of the reactions $e^{-} \, e^{+} \rightarrow e^{-} \, e^{+}$ and $e^{-} \, e^{+} \rightarrow \gamma \, \gamma$ at $29$ GeV}, Phys. Rev. D \textbf{34}, 3286 (1986).

\bibitem{Levi} M.E. Levi {\it et al.}, {\it Weak Neutral Currents in $e^+ e^-$ Collisions at $\sqrt{s} = 29$~GeV}, Phys. Rev. Lett. {\bf 51}, 1941 (1983).

\bibitem{Abrams1} G. S. Abrams {\it et al.}, {\it Measurement of Z decays into lepton pairs}, Phys. Rev. Lett. {\bf 63}, 2780 (1989).

\bibitem{Abrams2} G. S. Abrams {\it et al.}, {\it Measurements of Z-boson resonance parameters in $e^+ e^-$ annihilation}, Phys. Rev. Lett. {\bf 63}, 2173 (1989).

\bibitem{PDG} R.I. Workman {\it et al.}, Particle Data Group, Progr. Theor. Exp. Phys. p. 083C01 (2022) and 2023 update.

\bibitem{Chanowitz_1} M.S. Chanowitz, M.A. Furman, I. Hinchliffe, {\it Weak Interactions of Ultraheavy Fermions}, Phys. Lett. B {\bf 78}, 285 (1978).

\bibitem{Chanowitz_2} M.S. Chanowitz, M.A. Furman, I. Hinchliffe, {\it Weak Interactions of Ultraheavy Fermions (II)}, Nucl. Phys. B {\bf 153}, 402 (1979).

\bibitem{Wick} M. Jacob, G. C. Wick, {\it On the general theory of collisions for particles with spin}, Annals Phys. {\bf 7}, 404 (1959).


\bibitem{Logan} H. E. Logan, {\it Lectures on perturbative unitarity and decoupling in Higgs physics}, arXiv:hep-ph/2207.01064.

\bibitem{Wigner} D.A. Varshalovich, A.N. Moskalev, V.K. Khersonskii, {\it Quantum Theory of Angular Momentum: Irreducible Tensors, Spherical Harmonics, Vector Coupling Coefficients, 3nj Symbols}, World Scientific Publishing Company (1988).

\bibitem{Luzio} L. Di Luzio, J. F. Kamenik, M. Nardecchia, {\it Implications of perturbative unitarity for scalar di-boson resonance searches at LHC}, Eur. Phys. J. C {\bf 77}, 30 (2017). 

\bibitem{Chano_WW} M. S. Chanowitz, {\it Strong WW scattering at the end of the 90’s: Theory and experimental prospects}, pulished in Zuoz 1998, ``Hidden symmetries and Higgs phenomena", p. 81; arXiv:hep-ph/9812215.

\bibitem{SM_unit_1} B. W. Lee, C. Quigg, H. B. Thacker, {\it Weak Interactions at Very High-Energies: The Role of the Higgs Boson Mass}, Phys. Rev. D {\bf 16}, 1519 (1977).

\bibitem{SM_unit_2} B. W. Lee, C. Quigg, H. B. Thacker, {\it Strength of Weak Interactions at Very High Energies and the Higgs Boson Mass}, Phys. Rev. Lett. {\bf 38}, 883 (1977).

\bibitem{Bij_Higgs} J.J. Van Der Bij, {\it Two-loop large higgs mass correction to vector boson masses}, Nucl. Phys. B {\bf 248}, 141 (1984).

\bibitem{djouadi} A. Djouadi, {\it The Anatomy of electro-weak symmetry breaking. I: The Higgs boson in the standard model}, Phys. Rept. {\bf 457}, 1 (2008).






\bibitem{Pedro_LSV} G. P. de Brito, J. T. Guaitolini Junior, D. Kroff, P. C. Malta, C. Marques, {\it Lorentz violation in simple QED processes}, Phys. Rev. D {\bf 94}, 056005 (2016).

\bibitem{Dutta} S. Dutta, P. Konar, B. Mukhopadhyaya, S. Raychaudhuri, {\it Bhabha scattering with radiated gravitons at linear colliders}, Phys. Rev. D {\bf 68}, 095005 (2003).

\bibitem{Bufalo} R. Bufalo, {\it On the Bhabha scattering for $z=2$ Lifshitz QED},  Int. J. Mod. Phys. A {\bf 30}, 1550086 (2015).

\bibitem{Bufalo2} R. Bufalo, B.M. Pimentel, D.E. Soto, {\it Causal approach for the electron-positron scattering in Generalized Quantum Electrodynamics}, Phys. Rev. D {\bf 90}, 085012 (2014).

\bibitem{OPAL} OPAL Collaboration, {\it Test of the four-fermion contact interaction in $e+ e^-$ collisions at 130-40 GeV}, Phys. Lett. B {\bf 387}, 432 (1996).

\bibitem{ILC} The International Linear Collider. URL: \url{http://www.linearcollider.org/cms/}.

\bibitem{Bechtle} P. Bechtle, S. Heinemeyer, J. List, G. Moortgat-Pick, G. Weiglein, {\it Physics case for an $e^+ e^-$ collider at 500 GeV and above}, arXiv:hep-ph/2410.16191.







\bibitem{FCC} FCC collaboration, A. Abada {\it et al.}, {\it FCC-ee: The Lepton Collider: Future Circular Collider Conceptual Design Report Volume 2}, Eur. Phys. J. ST {\bf 228}, 261 (2019).

\bibitem{CEPC} CEPC Study Group collaboration, M. Dong {\it et al.}, {\it CEPC Conceptual Design Report: Volume 2 - Physics \& Detector}, arXiv:hep-ex/1811.10545.

\bibitem{CCC} C. Vernieri {\it et al.}, {\it Strategy for Understanding the Higgs Physics: The Cool Copper Collider}, JINST {\bf 18}, P07053 (2023).

\bibitem{CLIC} CLIC, CLICdp collaboration, {\it The Compact Linear $e^+ e^-$ Collider (CLIC): Physics Potential}, arXiv:hep-ex/1812.07986.

\bibitem{muon_c} C. Accettura {\it et al.}, {\it Towards a muon collider}, Eur. Phys. J. C {\bf 83}, 864 (2023).

\bibitem{CSKR_SUSY} C.N. Ferreira, J.A. Helay\"{e}l-Neto, M.B.D.S.M. Porto, {\it Cosmic string configuration in the supersymmetric CSKR theory}, Nucl. Phys. B {\bf 620} (2002) 181; arXiv: hep-th/0101008v2.

\bibitem{Denis} D. Cocuroci, M.J. Neves, L.P.R. Ospedal and J.A. Helay\"{e}l-Neto, {\it A 3-form Gauge Potential in 5D in connection with a Possible Dark Sector of 4D-Electrodynamics}, Eur. Phys. J. C {\bf 75} 7, 322 (2015).




\bibitem{BF} E.Harikumar, M. Sivakumar, {\it Duality and Massive Gauge Invariant Theories}, Phys. Rev. D {\bf 57}, 3794 (1998).

\bibitem{Nascimento:2014owa}
J.~R.~Nascimento, A.~Y.~Petrov, C.~Wotzasek and C.~A.~D.~Zarro,
\textit{``Three-dimensional Lorentz-violating action''},
Phys. Rev. D \textbf{89}, no.6, 065030 (2014).






\bibitem{proj_proca_1} R.J. Rivers, {\it Lagrangian theory for neutral massive spin-2 fields}, Nuovo Cimento, {\bf 34}, 387 (1964).

\bibitem{proj_proca_2} R. Kuhfuss, J. Nitsch, {\it Propagating modes in gauge field theories of gravity}, Gen. Rel. Grav. {\bf 18}, 1207 (1986).























\bibitem{Ints} G.S. Adkins, {\it Three-dimensional Fourier transforms, integrals of spherical Bessel functions, and novel delta function identities}, arXiv:math-ph/1302.1830v1.


\end{thebibliography}
\end{document}